\documentclass[twocolumn]{aastex61}

\usepackage{color}
\usepackage{epsfig}
\usepackage{amsmath}
\usepackage{enumerate}
\usepackage{epstopdf}
\usepackage{verbatim}
\usepackage{courier}
\usepackage{tabularx}
\usepackage{longtable}
\usepackage{gensymb}
\usepackage{threeparttablex}
\usepackage{booktabs} 
\usepackage{natbib}

\begin{document}

\title{Statistics of coronal dimmings associated with coronal mass ejections. I. Characteristic dimming properties and flare association}
\correspondingauthor{K. Dissauer}
\email{karin.dissauer@uni-graz.at}

\author{K. Dissauer}
\affiliation{Institute of Physics, University of Graz, A-8010 Graz, Austria}
\author{A. M. Veronig}
\affiliation{Institute of Physics, University of Graz, A-8010 Graz, Austria}
\affiliation{Kanzelh\"ohe Observatory, University of Graz, A-9521 Treffen, Austria}
\author{M. Temmer}
\affiliation{Institute of Physics, University of Graz, A-8010 Graz, Austria}
\author{T. Podladchikova}
\affiliation{Skolkovo Institute of Science and Technology, 143026 Moscow, Russia}
\author{K. Vanninathan}
\affiliation{Institute of Physics, University of Graz, A-8010 Graz, Austria}

\begin{abstract}
Coronal dimmings, localized regions of reduced emission in the EUV and soft X-rays, are interpreted as density depletions due to mass loss during the CME expansion. They contain crucial information on the early evolution of CMEs low in the corona. For 62 dimming events, characteristic parameters are derived, statistically analyzed and compared with basic flare quantities. 
On average, coronal dimmings have a size of \mbox{$2.15\times10^{10}$~km$^{2}$}, contain a total unsigned magnetic flux of $1.75\times10^{21}$~Mx, and show a total brightness decrease of $-1.91\times10^{6}$~DN, which results in a relative decrease of $\sim$60\% compared to the pre-eruption intensity level. Their main evacuation phase lasts for $\sim$50~minutes. 
The dimming area, the total dimming brightness, and the total unsigned magnetic flux show the highest correlation with the flare SXR fluence ($c\gtrsim0.7$). Their corresponding time derivatives, describing the dimming dynamics, strongly correlate with the GOES flare class ($c\gtrsim 0.6$). 
For 60\% of the events we identified core dimmings, i.e.~signatures of an erupting flux rope. They contain 20\% of the magnetic flux covering only 5\% of the total dimming area. Secondary dimmings map overlying fields that are stretched during the eruption and closed down by magnetic reconnection, thus adding flux to the erupting flux rope via magnetic reconnection.
This interpretation is supported by the strong correlation between the magnetic fluxes of secondary dimmings and flare reconnection fluxes ($c=0.63\pm0.08$), the balance between positive and negative magnetic fluxes ($c=0.83\pm0.04$) within the total dimmings and the fact that for strong flares ($>$M1.0) the reconnection and secondary dimming fluxes are roughly equal.
\end{abstract}

\section{Introduction} \label{sec:intro}
Coronal dimmings present distinct low coronal signatures of coronal mass ejections (CMEs) and contain crucial information on their initiation, early evolution and plasma properties. They appear as transient regions of strongly reduced emission at EUV \citep{Thompson:1998, Thompson:2000} and soft X-ray (SXR; \citealt{Hudson:1996,Sterling:1997}) wavelengths during the early CME evolution, and are explained by the density depletion caused by the evacuation of plasma during the initial CME expansion. This interpretation is supported by the simultaneous and co-spatial observations of coronal dimmings in different wavelengths \citep{Zarro:1999}, spectroscopic observations showing plasma outflows in dimming regions \citep{Harra:2001,Tian:2012}, as well as DEM studies indicating localized density drops in dimming regions up to 70\% \citep{Vanninathan:2018}. 

Two different types of dimmings are in general distinguished \citep{Mandrini:2005,Mandrini:2007}. Core dimmings mark the footpoints of the evacuated flux rope and are observed as localized regions, close to the eruption site in opposite magnetic polarity regions \citep{Hudson:1996,Webb:2000}. Due to their proximity to the flare site and the line-of-sight integrated emission in the corona their full extent is hard to identify, and the number of events where both footpoints of the flux rope appear clearly pronounced as bipolar dimmings is limited \citep[e.g.][]{Attrill:2006,Dissauer:2016}. Secondary dimmings are regions of reduced emission due to the expansion of the CME body and overlying fields that are erupting and therefore observed as widespread and more shallow dimming regions \citep{Attrill:2007,Mandrini:2007}. They are a reflection of the expanding CME in the low corona.

There have been a few attempts to establish a dimming/CME mass relationship \citep{Harrison:2000,Harrison:2003,Aschwanden:2009,Aschwanden:2016,Mason:2016}. However, to date only few statistical studies on coronal dimmings and their relationship to CMEs exist \citep{Reinard:2008, Bewsher:2008,Aschwanden:2016,Mason:2016,Krista:2017}. Especially statistical studies analyzing the properties of coronal dimmings in detail are rare \citep{Reinard:2008,Aschwanden:2016,Mason:2016,Krista:2017}.

We present a statistical analysis of coronal dimmings and their corresponding CMEs where the parameters of both phenomena can be simultaneously derived with good accuracy using multi-spacecraft observations.
In \cite{Dissauer:2018}, we introduced a new method for the detection of coronal dimmings, which allows next to the extraction of core dimming regions also the detection of the more shallow and widespread secondary dimming regions. To determine the physical properties of coronal dimmings, characteristic parameters are introduced that describe their dynamics, morphology, magnetic properties and their brightness evolution. 
This method is applied to a statistical set of 62 coronal dimming events in optimized multi-point observations, where the dimming is observed against the solar disk by SDO/AIA and the associated CME evolution close to the limb by STEREO/EUVI, COR1 and COR2. This paper is the first in a series of two statistical papers on coronal dimmings. Here, we present the detailed statistical analysis of coronal dimmings and compare their properties to basic parameters of their associated flares.
The second paper will focus on the relationship between characteristic coronal dimming and CME parameters, like mass, speed and acceleration characteristics.
\section{Data set}
\subsection{Event selection}
Dimming events were selected such that they occurred on-disk for SDO, while the associated CME was observed close to the limb by at least one of the twin STEREO spacecraft. In this way we make use of the optimum combination of simultaneous multi-point observations for Earth-directed events with minimal projection effects. We study the time range between May 2010, when SDO science data start and when the longitudinal separation of the STEREO-A and -B s/c relative to the Sun-Earth line are $\pm75\degree$, and September 2012 when the separation to the Sun-Earth-line has increased to about $\pm125\degree$.

For the event selection, we used two approaches. On the one hand, we select all events that occurred within the considered time range in the SDO/AIA catalog of large-scale EUV waves\footnote{http://www.lmsal.com/nitta/movies/AIA\_Waves} described in \cite{Nitta:2013}. On the other hand, we identified all halo CMEs in the CDAW LASCO/CME catalog\footnote{https://cdaw.gsfc.nasa.gov/CME\_list/} described in \cite{Yashiro:2004}, as those represent CMEs that are likely moving in the Sun-observer line. Both phenomena, i.e.~halo CMEs and EUV waves, are known to be often associated with coronal dimmings.
We focus on events where the eruption site lies within $\pm40\degree$ from the central meridian of the Sun.
For the second approach, we visually checked the STEREO-A and STEREO-B COR1 movies to investigate whether the selected halo CMEs were front-sided or not.
To this aim, we linked the GOES flare list\footnote{http://www.ngdc.noaa.gov/stp/satellite/goes/index.html} to the remaining halo CMEs. In order to meet the same criteria as for the EUV waves, we again select only events where the eruption site lies within $\pm40\degree$. 

We cross-check the halo CME events with the EUV wave events identified in order to avoid double entries. By merging the results of the two data sets, we arrive at a total of 76 suitable candidate events. Five events were associated with large-scale filament eruptions. The evacuation of cool filament material results in a darkening in EUV filters that is not associated with the coronal dimming that we aim to detect. Therefore, these events were withdrawn from the statistical sample. Further nine events showed no clear dimming formation and were therefore also excluded. The final data set covers 62 coronal dimming events consisting of the following subsets: 23 halo CMEs\footnote{halo CMEs as identified by SOHO/LASCO} (19 of them are associated to an EUV wave) and 39 EUV waves that are not associated to a halo CME. Or alternatively expressed: 58 EUV waves (19 of them are associated to a halo CME) and 4 halo CMEs that are not associated to an EUV wave. The distribution among the GOES classes of the associated flares is, B: 7, C: 25, M: 26, X: 4.

\begin{figure}
\centering
\includegraphics[width=1.0\columnwidth]{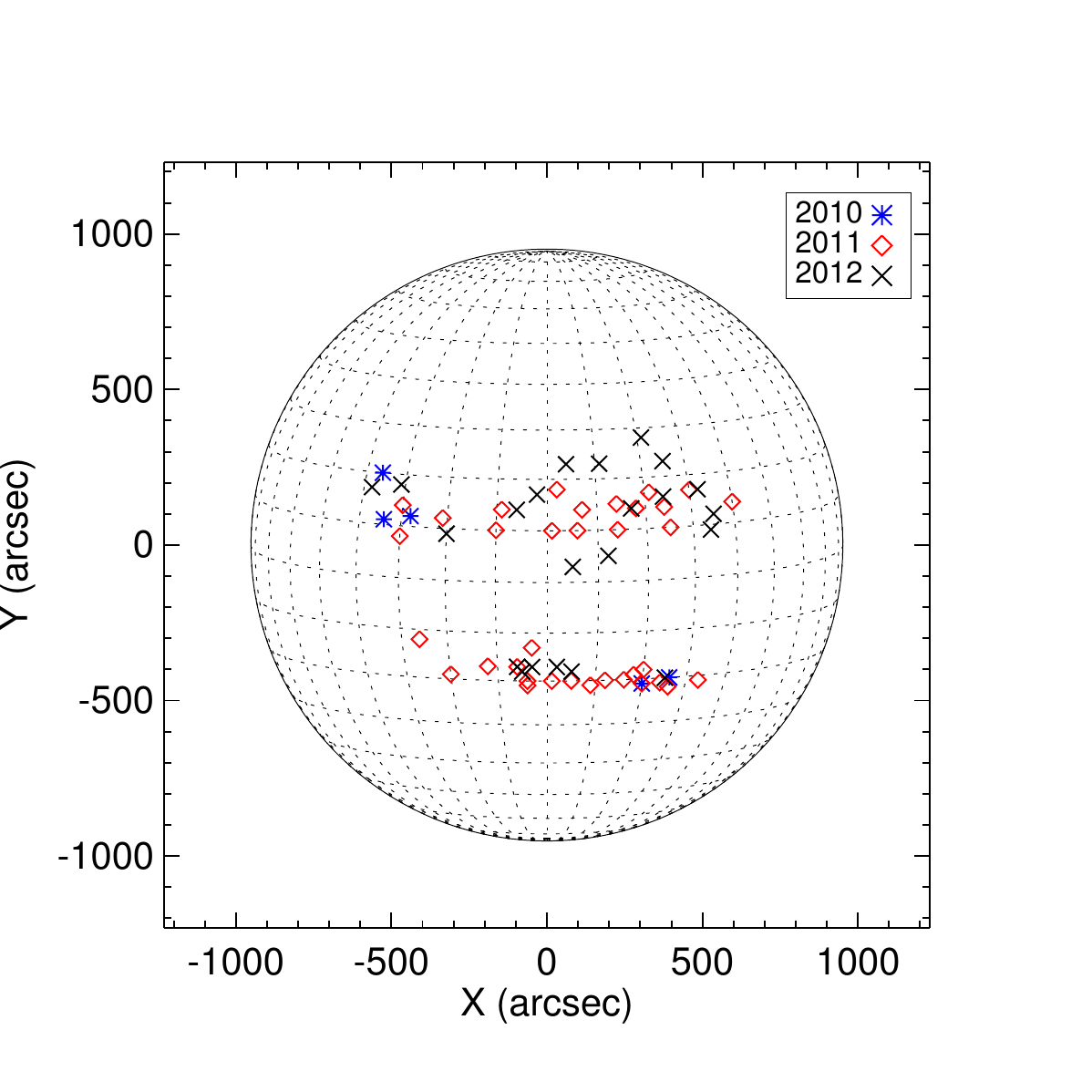}
\caption{Position of the source regions of the coronal dimming events under study as observed by SDO's point of view. Different colors and symbols represent the different years of occurrence.}
\label{fig:position}
\end{figure}

The main characteristics of the selected events including information on the source active region and their associated flares are given in Table~\ref{tab:events}. Figure~\ref{fig:position} shows the position of their source region on the solar disk.

\renewcommand{\arraystretch}{0.8}
\begin{ThreePartTable}
\begin{TableNotes}
\item \textbf{Note.} We list the date, start, peak and end time of the associated flare (from the GOES flare catalog, or derived from the GOES SXR flux using the same criteria as in the GOES flare catalog), the NOAA Active Region number, its heliographic position, the peak of the GOES SXR flux $F_{P}$, the maximum of its derivative $\dot{F}_{P}$, and the SXR flare fluence $F_{T}$. $A_{\text{rbn}}$ and $\Phi_{\text{rbn}}$ are the flare ribbon area and the total unsigned reconnected flux extracted from flare ribbon observations by \cite{Kazachenko:2017}. The second column indicates whether the event was associated with an EUV wave or not (W/--) and whether the associated CME was identified as halo CME by SOHO/LASCO (halo).
\end{TableNotes}
\begin{longtable*}{p{0.5cm}llp{0.7cm}p{0.7cm}p{0.7cm}p{0.9cm}p{1.4cm}p{1.3cm}p{1.6cm}p{1.3cm}p{0.8cm}p{0.8cm}}
\caption{Overview of the events under study}\label{tab:events} \\
\toprule
\#&Catalog&Date&Start Time [UT]&Peak Time [UT]&End Time [UT]&NOAA AR&Flare Location&$F_{P}$ [W~m$^{-2}$]&$\dot{F}_{P}$ [W~m$^{-2}$~s$^{-1}$]&$F_{T}$ [J~m$^{-2}$] &$A_{\text{rbn}}$ [km$^{2}$] ($10^{18}$)&$\Phi_{\text{rbn}}$ [Mx] ($10^{21}$) \\ \hline
1&W&20100716&15:13&15:50&16:02&-&S21 W20&1.82E-07&2.23E-10&-&-&- \\
2&W (halo)&20100801&07:37&08:56&10:01&11092&N20 E36&3.24E-06&1.26E-09&1.50E-02&4.72&2.96 \\
3&W (halo)&20100807&17:54&18:24&18:46&11093&N11 E34&1.04E-05&1.28E-08&1.80E-02&9.50&4.75 \\
4&W&20101016&19:06&19:12&19:14&11112&S20 W26&3.15E-05&1.58E-07&6.40E-03&5.30&2.76 \\
5&W&20101111&18:52&19:01&19:05&11124&N12 E28&9.26E-07&2.63E-09&4.20E-04&-&- \\
6&W (halo)&20110213&17:28&17:37&17:46&11158&S20 E04&6.68E-05&2.82E-07&4.00E-02&6.90&5.12 \\
7&W&20110214&02:34&02:41&02:45&11158&S21 E04&1.68E-06&3.66E-09&7.10E-04&1.80&0.76 \\
8&W&20110214&04:28&04:48&05:08&11158&S20 W01&8.34E-06&1.35E-08&1.30E-02&2.93&2.48 \\
9&W (halo)&20110214&17:20&17:25&17:32&11158&S20 W05&2.24E-05&1.08E-07&8.70E-03&-&- \\
10&W (halo)&20110215&01:44&01:56&02:05&11158&S20 W12&2.31E-04&6.71E-07&1.60E-01&15.40&11.60 \\
11&W&20110215&04:27&04:32&04:36&11158&S21 W09&5.30E-06&1.49E-08&1.90E-03&-&- \\
12&W&20110215&14:32&14:43&14:50&11158&S20 W16&4.88E-06&1.22E-08&3.40E-03&2.73&1.64 \\
13&W&20110307&13:44&14:30&14:56&11166&N12 E21&1.99E-05&2.04E-08&6.20E-02&12.50&5.18 \\
14&W&20110308&18:52&18:56&18:58&11166&N10 W01&1.00E-07&8.46E-09&-&-&- \\
15&W&20110325&23:08&23:21&23:29&11176&S12 E26&1.02E-05&2.89E-08&8.00E-03&4.37&1.56 \\
16&W (halo)&20110602&07:21&07:46&07:56&11227&S19 E20&3.78E-06&4.58E-09&5.10E-03&5.16&1.70 \\
17&-- (halo)&20110621&01:21&03:26&04:20&11236&N14 W09&7.75E-06&3.51E-09&4.30E-02&3.28&1.13 \\
18&W&20110703&00:00&00:23&00:32&11244&N14 W24&9.54E-07&1.86E-09&1.30E-03&-&- \\
19&W&20110711&10:46&11:03&11:10&11249&S17 E06&2.63E-06&3.58E-09&2.40E-03&0.83&0.26 \\
20&W&20110802&05:58&06:18&06:41&11261&N15 W14&1.49E-05&1.27E-08&3.90E-02&10.90&7.07 \\
21&W (halo)&20110803&13:17&13:47&14:09&11261&N17 W30&6.08E-05&7.46E-08&1.20E-01&11.10&7.61 \\
22&W (halo)&20110906&01:35&01:49&02:04&11283&N14 W07&5.38E-05&1.67E-07&5.40E-02&7.21&3.26 \\
23&W (halo)&20110906&22:12&22:20&22:23&11283&N14 W18&2.14E-04&1.01E-06&5.80E-02&13.10&5.92 \\
24&W&20110908&15:32&15:45&15:52&11283&N14 W40&6.75E-05&1.89E-07&4.20E-02&15.70&7.33 \\
25&W&20110926&14:36&14:46&15:02&11302&N14 E30&2.62E-05&8.74E-08&2.60E-02&12.00&6.42 \\
26&W&20110927&20:43&20:58&21:11&11302&N14 E09&6.44E-06&9.29E-09&7.40E-03&5.09&1.97 \\
27&W&20110930&03:36&03:59&04:12&11305&N10 E10&7.73E-06&1.51E-08&8.60E-03&4.19&1.89 \\
28&W&20111001&09:20&09:59&10:15&11305&N10 W06&1.28E-05&1.24E-08&2.90E-02&8.85&3.60 \\
29&W&20111002&00:37&00:49&00:59&11305&N10 W14&3.92E-05&1.13E-07&2.80E-02&5.34&2.42 \\
30&W&20111002&21:20&21:48&22:04&11305&N10 W25&7.65E-06&2.53E-08&9.10E-03&5.98&2.63 \\
31&W&20111010&14:29&14:34&14:36&11313&S13 E03&4.82E-06&2.30E-08&6.90E-04&0.94&0.32 \\
32&W&20111115&-&-&-&11347&N08 E30&-&-&-&-&- \\
33&W&20111124&23:56&00:41&01:32&11354&S18 W20&1.50E-06&1.31E-09&-&-&- \\
34&W&20111213&03:07&03:11&03:15&11374&S17 E12&8.14E-07&1.05E-09&-&-&- \\
35&W&20111222&01:56&02:08&02:20&11381&S19 W18&5.48E-06&1.01E-08&5.50E-03&2.31&1.59 \\
36&W&20111225&08:49&08:55&09:00&11387&S21 W20&5.57E-06&8.32E-09&3.10E-03&3.63&1.83 \\
37&W&20111225&18:11&18:16&18:19&11387&S22 W26&4.14E-05&1.97E-07&1.10E-02&7.32&4.45 \\
38&W&20111225&20:23&20:28&20:32&11387&S21 W24&8.02E-06&3.53E-08&2.50E-03&2.59&1.43 \\
39&W&20111226&02:13&02:26&02:35&11387&S21 W33&1.52E-05&3.59E-08&1.20E-02&3.82&2.55 \\
40&W&20111226&11:22&11:47&12:18&11384&N18 W02&5.76E-06&5.25E-09&1.40E-02&3.95&1.09 \\
41&W (halo)&20120123&03:38&03:58&04:34&11402&N28 W21&8.76E-05&1.17E-07&2.00E-01&27.20&17.20 \\
42&W (halo)&20120307&00:02&00:24&00:40&11429&N18 E31&5.43E-04&1.11E-06&6.70E-01&35.20&30.40 \\
43&W (halo)&20120309&03:21&03:53&04:18&11429&N17 E02&6.36E-05&8.40E-08&1.30E-01&23.00&14.50 \\
44&W (halo)&20120310&17:15&17:44&18:29&11429&N16 W24&8.49E-05&1.88E-07&2.60E-01&27.10&16.90 \\
45&W&20120314&15:07&15:21&15:36&11432&N14 E06&2.82E-05&7.97E-08&2.90E-02&7.02&3.09 \\
46&W&20120317&20:32&20:39&20:41&11434&S20 W25&1.37E-05&6.61E-08&3.60E-03&2.59&1.32 \\
47&W (halo)&20120405&20:49&21:09&21:56&11450&N17 W32&1.59E-06&1.56E-09&5.10E-03&2.98&1.28 \\
48&W (halo)&20120423&17:37&17:50&18:04&11461&N14 W17&2.08E-06&2.94E-09&2.50E-03&1.95&0.60 \\
49&-- (halo)&20120511&23:02&23:43&00:27&11476&N05 W12&3.24E-06&2.11E-09&1.30E-02&6.14&3.17 \\
50&W&20120603&17:48&17:54&17:57&11496&N17 E38&3.44E-05&1.64E-07&7.00E-03&-&- \\
51&W&20120606&19:53&20:06&20:12&11494&S18 W05&2.19E-05&6.15E-08&1.30E-02&4.98&2.05 \\
52&-- (halo)&20120614&12:51&14:35&15:56&11504&S17 E06&1.92E-05&9.67E-09&1.20E-01&4.54&3.88 \\
53&W&20120702&10:43&10:51&10:56&11515&S18 E05&5.61E-05&2.10E-07&2.70E-02&10.60&3.84 \\
54&W&20120702&19:59&20:06&20:12&11515&S17 E03&3.80E-05&1.49E-07&1.80E-02&10.50&4.78 \\
55&W (halo)&20120704&16:33&16:39&16:48&11513&N12 W35&1.89E-05&6.62E-08&1.20E-02&8.47&3.56 \\
56&W (halo)&20120712&16:11&16:52&17:35&11520&S17 W02&1.42E-04&1.07E-07&4.60E-01&12.80&8.64 \\
57&W (halo)&20120813&12:33&12:40&12:48&11543&N23 W04&2.88E-06&1.08E-08&1.80E-03&2.95&1.16 \\
58&W (halo)&20120814&00:23&00:30&00:42&11543&N23 W11&3.50E-06&1.04E-08&-&2.54&1.04 \\
59&W&20120815&03:37&03:45&03:54&11543&N23 W25&8.48E-07&1.77E-09&5.80E-04&-&- \\
60&-- (halo)&20120902&01:49&01:58&02:10&11560&N03W05&2.99E-06&6.42E-09&2.80E-03&2.97&1.33 \\
61&W&20120925&04:24&04:35&04:48&11577&N09 E20&3.61E-06&9.25E-09&3.80E-03&1.41&0.42 \\
62&W&20120927&23:35&23:57&00:34&11575&N09 W34&3.76E-06&4.10E-09&9.40E-03&6.02&2.33 \\ 
\bottomrule
\insertTableNotes\\
\end{longtable*}
\end{ThreePartTable}

\subsection{Data and data reduction}
Coronal dimmings are analyzed using high-cadence data from seven different extreme-ultraviolet (EUV) filters of the Atmospheric Imaging Assembly (AIA; \citealt{Lemen:2012}) on-board the Solar Dynamics Observatory (SDO; \citealt{Pesnell:2012}) covering a temperature range of \mbox{$\approx 50.000- 1.0\times 10^{7}$~K}. To study their magnetic properties, the 720~s line-of-sight (LOS) magnetograms of the SDO/Helioseismic and Magnetic Imager (HMI; \citealt{Scherrer:2012,Schou:2012}) are used. 

The data is rebinned to \mbox{$2048 \times 2048$} pixels under the condition of flux conservation. Standard Solarsoft IDL software is used for data reduction (\mbox{\texttt{aia\_prep.pro}} and \mbox{\texttt{hmi\_prep.pro}}). Only AIA images where the automatic exposure control (AEC) was not activated are used. Each data set is rotated to a common reference time using \mbox{\texttt{drot\_map.pro}} to correct for differential rotation. Furthermore, the detection of coronal dimming regions is restricted to a subfield of \mbox{$1000\times 1000$~arcsecs} around the center of the eruption.
To study the evolution of coronal dimming events our time series covers 12 hours, starting 30~minutes before the associated flare. For the first two hours the full-cadence (12~s) observations of SDO/AIA are used, while for the remaining time series the cadence of the observations is successively reduced to 1, 5, and 10~minutes. We note that the main focus of this study lies on the initial, impulsive expansion phase of the dimming and not on its recovery phase later on, which is also captured within this time series.
\section{Methods and Analysis}
We statistically analyze coronal dimmings observed in seven different EUV channels of SDO/AIA.
In Section~\ref{sec:detection} and \ref{sec:parameters}, we summarize the main steps of the dimming detection algorithm and list characteristic parameters describing their physical properties based on \cite{Dissauer:2018}.
We use two sample events, 2011 June 21 (\#17, gradual C7.7 flare, associated with a halo CME but no EUV wave) and 2012 June 6 (\#51, impulsive M2.1 flare, associated with a partial halo CME and an EUV wave), to illustrate the method and to show the time evolution of selected dimming parameters in SDO/AIA 211~\AA.
In Section~\ref{sec:impulsive_phase}, we introduce the impulsive phase of the dimming in order to calculate dimming parameter values for the events under study. Section~\ref{sec:statistics} gives an overview on the statistical analysis performed and how error bars are calculated.

\subsection{Dimming detection}\label{sec:detection} 
To identify coronal dimming regions and to follow their evolution, a thresholding technique applied on logarithmic base-ratio images is used \citep{Dissauer:2018}. The base image represents the median over the first ten images within the time series in each pixel, i.e.~30~min before the start of the associated flare.
All pixels whose logarithmic (log10) ratio intensity decreased below $-0.19$, corresponding to a change of about 35\% in linear space, are identified as dimming pixels. In order to reduce noise and the number of misidentified pixels, morphological operators are used to smooth the extracted regions, i.e.~small features are removed, while small gaps are filled. Dimming regions are in general complex, and different parts may grow and recover on different timescales. From instantaneous dimming masks, representing all dimming pixels detected at a specific time step $t_{m}$, we calculate cumulative dimming pixel masks by combining all dimming pixels identified \textit{up to} the time $t_{m}$. In this way, we are able to capture the full extent of the total dimming region over time.

In addition we also identify core dimming regions. Since they mark the footpoints of the erupting flux rope during the CME expansion, they should be observed close to the eruption site in regions of opposite magnetic polarity and reveal a strong intensity decrease. We aim to detect core dimmings as a subset of the total dimming region using minimum intensity maps. These maps represent the minimum intensity of each identified dimming pixel individually over the considered time range, calculated for base-difference and logarithmic base-ratio data. 
Core dimming pixels are then defined to be the darkest pixels in terms of absolute and relative intensity change in minimum intensity maps during the early phase of the dimming expansion. For further details on the method we refer to \cite{Dissauer:2018}.
\begin{figure*}
\centering
\includegraphics[width=0.8\textwidth]{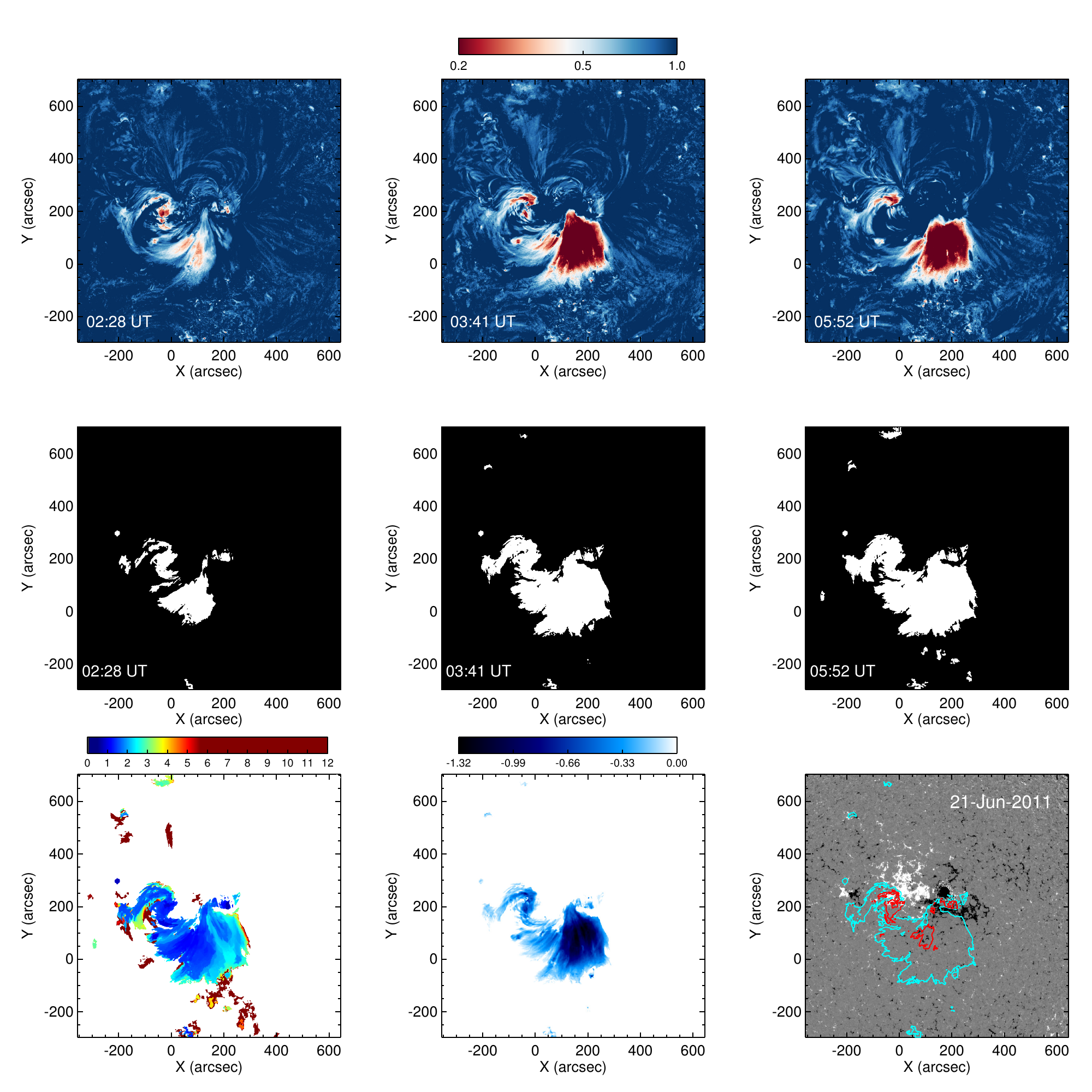}
\caption{Evolution of the coronal dimming region of event \#17 on 2011 June 21. Time sequence of SDO/AIA 211\AA~images during the dimming expansion (top row) together with the corresponding cumulative dimming pixel masks (middle row). Bottom right: Corresponding timing maps of the coronal dimming region indicating the time at which a pixel is detected as a dimming pixel for the first time (color coded in hours after the flare onset). Bottom middle: Minimum intensity map of logarithmic base-ratio data. Bottom right: SDO/HMI LOS magnetogram with the total dimming region identified during the impulsive phase (cyan contours) and potential core dimming regions (red contours) overplotted.}
\label{fig:detection_20110621_0052}
\end{figure*}
\begin{figure*}
\centering
\includegraphics[width=0.8\textwidth]{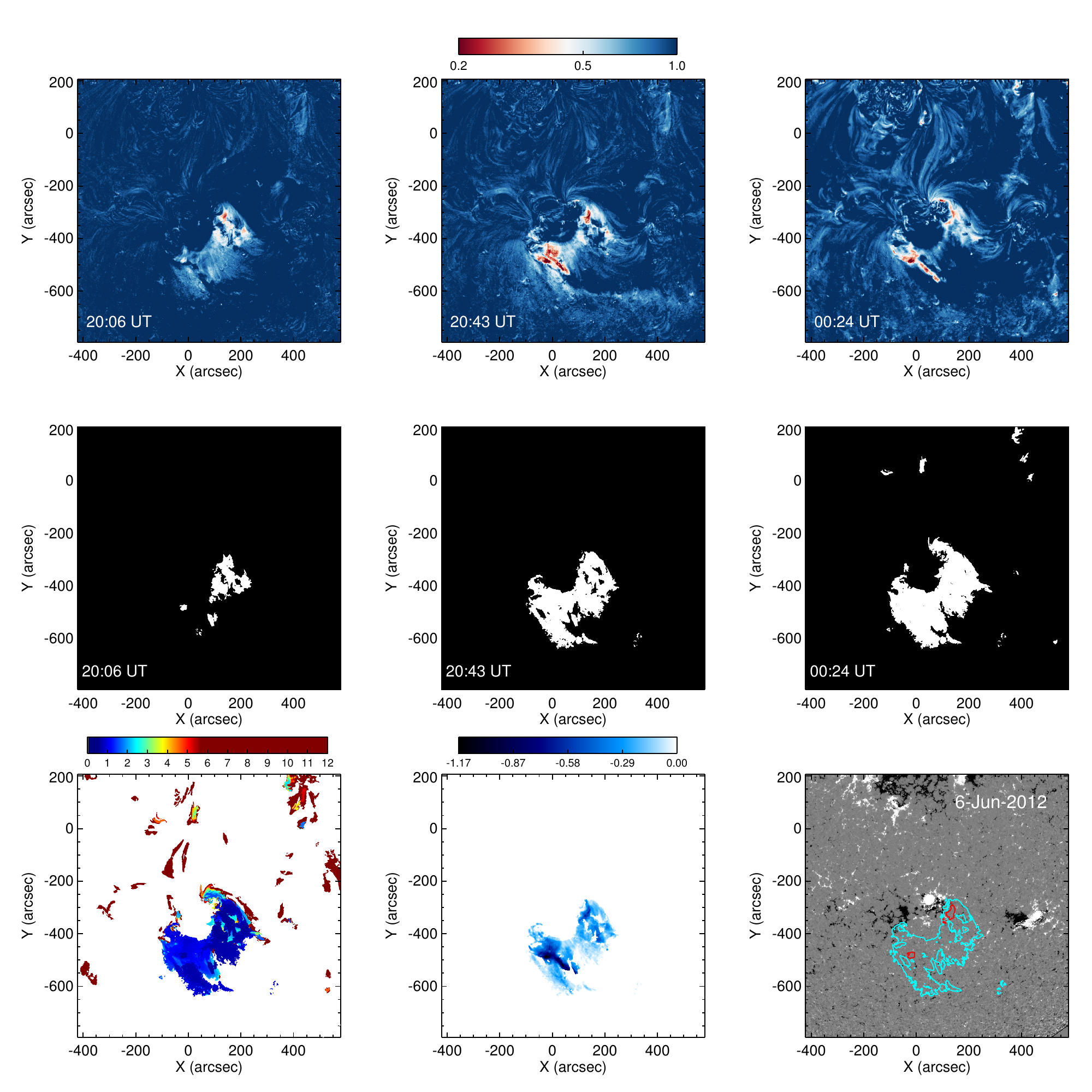}
\caption{Same as in Figure~\ref{fig:detection_20110621_0052} but for event \#51 on 2012 June 6.}
\label{fig:detection_20120606_1924}
\end{figure*}

Figure~\ref{fig:detection_20110621_0052}~and \ref{fig:detection_20120606_1924}~give an overview of the dimming detection of the sample events \#17 and \#51. Panels in the top row show a sequence of logarithmic base-ratio images of SDO/AIA 211 \AA~ filtergrams indicative of the time evolution of the dimming region. Regions of moderate and strong intensity decrease (i.e.~coronal dimmings) appear from light blue to red, while regions where the intensity did not change or increased appear dark blue. Panels in the middle row represent the corresponding cumulative dimming pixel masks and show all identified dimming pixels up to the given time step. The bottom left panel shows the final cumulative dimming pixel mask for the full time range of 12 hours, where each pixel is color coded by the time of its first detection (in hours after the flare onset). In the bottom middle panel the corresponding minimum intensity map of logarithmic base-ratio data is plotted, showing for each pixel identified during the impulsive phase (see Sect.~\ref{sec:impulsive_phase}) its minimum intensity. The corresponding SDO/HMI LOS magnetogram is given in the bottom right, illustrating the position and extent of the total dimming region identified during the impulsive phase in cyan contours. The red contours mark the detected potential core dimming regions.
\subsection{Characteristic dimming parameters}\label{sec:parameters}
For each event in our sample, we extract characteristic parameters describing the physical properties of the coronal dimmings.

To determine the size of the dimming regions, we cumulate the area of newly detected dimming pixels over time. Thus, $A(t_{n})$ represents the area of all dimming pixels that are detected until $t_{n}$, the end of the dimming evolution. The area growth rate $\dot{A}(t_{i})$, i.e.~how fast the dimming region is growing at time $t_{i}<t_{n}$, is calculated as the time derivative of the area evolution. 
In addition, also the ``magnetic area" of the dimming $A_{\Phi}(t_{n})$, i.e.~the area where the magnetic flux density $B$ obtained from SDO/HMI LOS magnetograms exceeds the noise level ($|B|>10$~G), and its derivative $\dot{A}_{\Phi}(t_{i})$ are extracted.

Like the area, also the magnetic flux is extracted cumulatively, as total unsigned $\Phi(t_{n})$, positive $\Phi_{+}(t_{n})$ and negative magnetic flux $\Phi_{-}(t_{n})$, in the ``magnetic area" of the dimming region, respectively. The corresponding magnetic flux rates $\dot{\Phi}(t_{i})$ and the mean unsigned magnetic flux density $\bar{B}_{\text{us}}(t_{i})$ are also calculated.

The brightness evolution of the dimming region is studied for a constant area $A(t_{n})$, represented by cumulative dimming pixel masks at the end of the dimming expansion $t_{n}$. $I_{\text{cu}}(t_{i})$ is defined as the sum of the intensities at any time step $t_{i}$ of all dimming pixels detected until $t_{n}$. In this way the change in the dimming brightness results only from the intensity change of dimming pixels over time and not from the changing dimming area. The brightness change rate is given by the corresponding time derivative $\dot{I}_{\text{cu}}(t_{i})$.
A measure for the total brightness of the dimming region is obtained by extracting the minimum in the time evolution of $I_{\text{cu}}$.
Coronal dimmings develop fast over time, this means that some parts may reach their lowest intensities before other regions of the dimming. Therefore, we also calculate the total brightness of the dimming region using minimum intensity maps. These maps include for all dimming pixels identified during the impulsive phase (see Section \ref{sec:impulsive_phase}) the minimum intensity of each pixel individually over time. $I_{\text{min}}$ is then derived as the sum of all pixel intensities in the minimum intensity map.
Note that all parameters describing the brightness of coronal dimmings are calculated from base-difference images, i.e.~they describe the decrease of the intensity in the dimming region with respect to the pre-event intensity level.
\begin{figure}
\centering
\includegraphics[width=1.0\columnwidth]{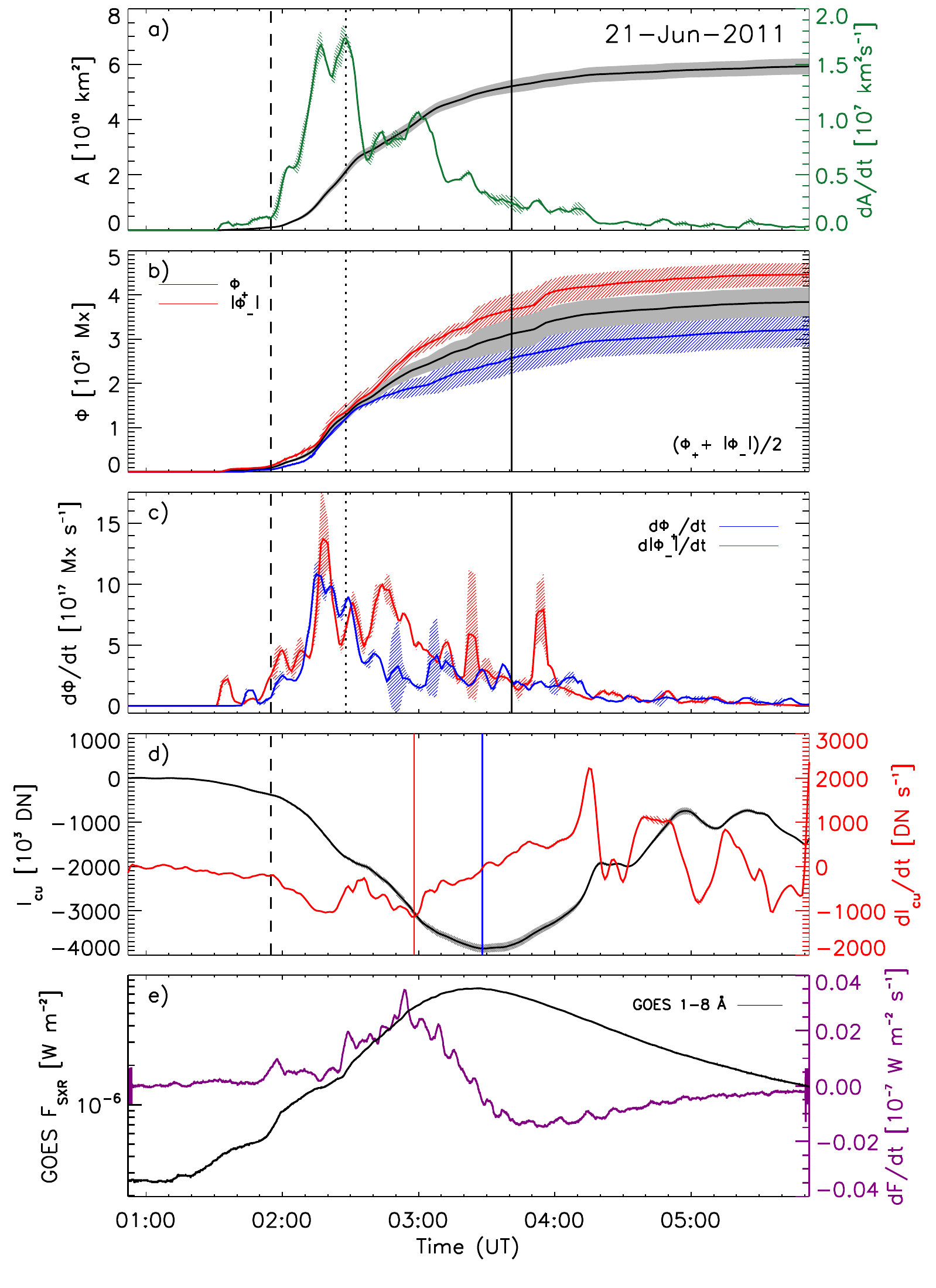}
\caption{Time evolution of selected coronal dimming parameters for the dimming event that occurred on 2011 June 21 (\#17). From top to bottom we plot (a) the dimming area $A$ (black line) and its time derivative, the area growth rate $\dot{A}$ (red line), (b) the positive (blue line), absolute negative (red line) and total unsigned magnetic flux (black line), (c) the corresponding magnetic flux rates $\dot{\Phi}_{+}$ (blue line), $\dot{\Phi}_{-}$ (red line), $\dot{\Phi}$ (black line), (d) the dimming brightness $I_{\text{cu}}$ (black line) and its time derivative, the brightness change rate $\dot{I}_{\text{cu}}$ (red line), and (e) the GOES 1.0-8.0 \AA~SXR flux and its derivative (purple). The shaded bands represent the 1$\sigma$ error bars for each parameter. The vertical dashed, dotted and solid lines mark the start, maximum and end of the impulsive phase of the dimming. The vertical blue and red lines indicate the minimum in the dimming brightness profiles.}
\label{fig:profiles_20110621}
\end{figure}
\begin{figure}
\includegraphics[width=1.0\columnwidth]{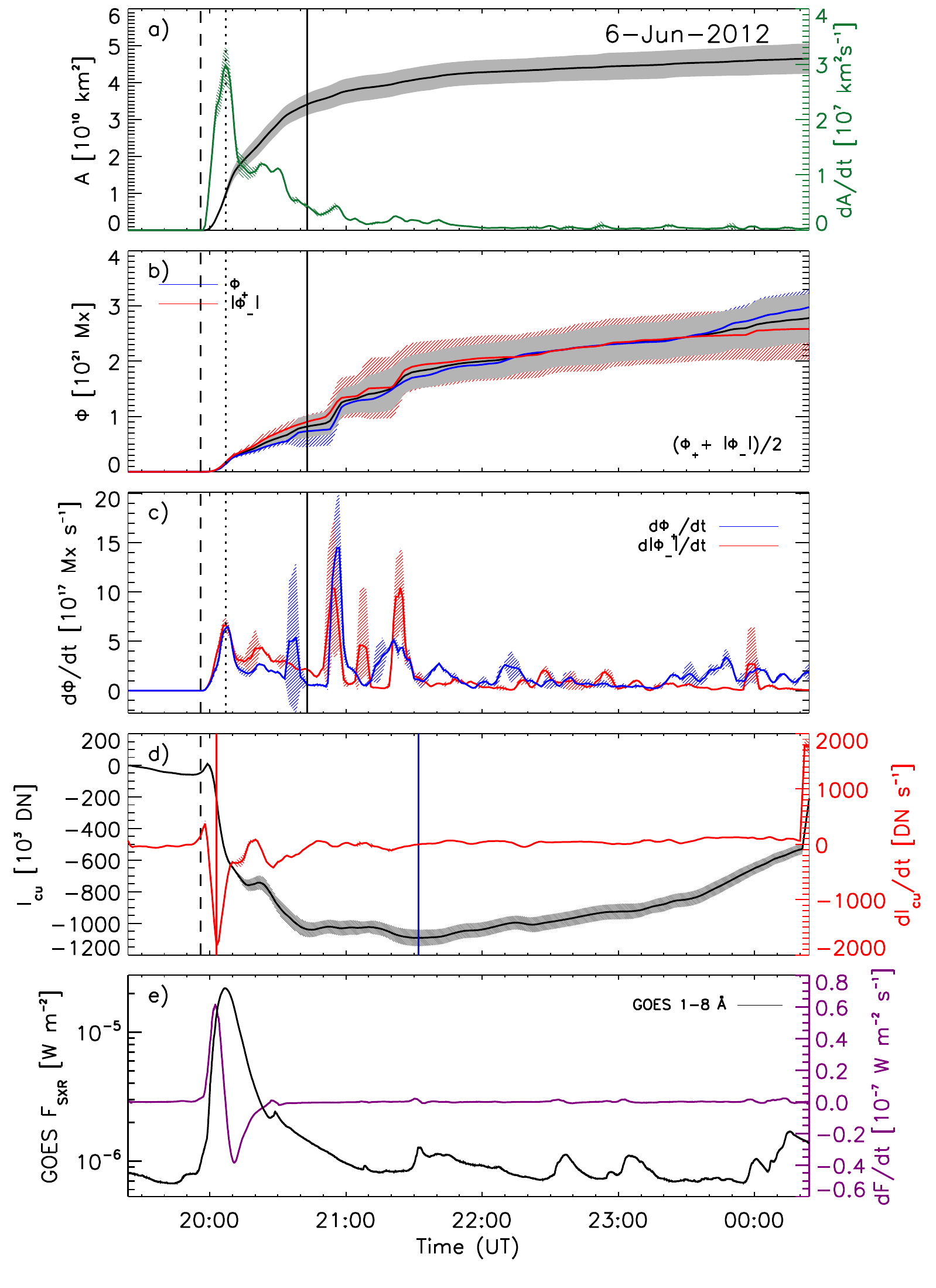}
\caption{Time evolution of coronal dimming parameters for dimming event \#51 that occurred for June 6, 2012. Same as Figure~\ref{fig:profiles_20110621}.}
\label{fig:profiles_20120606}
\end{figure}

Figure~\ref{fig:profiles_20110621} and \ref{fig:profiles_20120606} show the time evolution of selected dimming parameters (panels (a-d)) for the sample events \#17 and \#51, respectively. The solid, thick lines represent the mean, the shaded regions the 1$\sigma$ standard deviation used as error bars of each parameter (see Section~\ref{sec:statistics}).
In Panel~(a) the cumulative dimming area $A$ (black) and its time derivative, the area growth rate $\dot{A}$ (green) are plotted. Both events show a characteristic peak in the area growth rate profile, which is co-temporal with the rise in the GOES SXR flux of the corresponding flares (panel (f)). This peak in $\dot{A}(t)$ is also later used to define the impulsive phase of the dimming (see Section~\ref{sec:impulsive_phase}).
Panel~(b) shows the positive $\Phi_{+}$ (blue), absolute negative $|\Phi_{-}|$ (red) and total unsigned magnetic flux $\Phi$ (black) through the dimming region for each event. The amount of positive and negative flux is almost balanced. The corresponding magnetic flux change rates are plotted in panel (c). Panel (d) shows the time evolution of the dimming brightness $I_{\text{cu}}$ and its time derivative, the brightness change rate $\dot{I}_{\text{cu}}$. The blue and red vertical lines mark the minimum in each profile. Both events show a pronounced decrease in the dimming brightness profile. For \#17 it drops more gradually and increases soon after reaching its minimum; in contrast to \#51, which shows a rapid decrease to its minimum and remains at this level over a long duration.  

The introduced dimming parameters reflect the properties of the total dimming region (i.e.~including both core and secondary dimmings) and are obtained either as cumulative sums until the end of the impulsive phase, or in case of derivatives as the minimum/maximum value of the corresponding profile during the impulsive phase.
In order to obtain more information on the distribution of the different dimming types, we also calculate the area and the magnetic fluxes separately for the extracted core dimming regions.

\subsection{Impulsive phase of the dimming}\label{sec:impulsive_phase}
Coronal dimmings are regions of reduced emission that are growing while the associated CME is expanding. Therefore, the evolution of its area $A(t)$ and especially its area growth rate, $\dot{A}(t)$, can be used to determine the impulsive phase of the dimming evolution.
We define the impulsive phase of the dimming via the highest peak identified in its area growth rate profile, $\dot{A}_{\text{max}}$. 
The onset of the impulsive phase ($t_{\text{start}}$) is then defined as the local minimum that occurs closest in time before the peak.
The end of the impulsive phase ($t_{\text{end}}$) is defined as the time when the area growth rate $\dot{A}$ meets one of the two following criteria for the first time:
\begin{equation}
\dot{A}(t)\le\bar{\dot{A}}+3\sigma_{\dot{A}} \quad\lor \quad \dot{A}(t)\le 0.15\cdot\dot{A}_{\text{max}} \;,
\end{equation}
where $\bar{\dot{A}}$ and $\sigma_{\dot{A}}$ are the mean and standard deviation of the baseline, respectively.
The baseline parameters ($\bar{\dot{A}}$ and $\sigma_{\dot{A}}$) are calculated over a duration of 1~hour, starting 5~hrs after the start of the time series, where the main expansion of the dimming is already over and the number of newly detected dimming pixels is significantly decreased. During this time period secondary dimming regions in general replenish, while core dimming regions are still present \citep{Vanninathan:2018}. We did not choose the very end of the time range since effects of differential rotation become dominant and artifacts in the dimming detection due to local variations in the corona.
Panel (a) in Figure~\ref{fig:profiles_20110621} and \ref{fig:profiles_20120606} shows examples of the time evolution of the area growth rate for the sample events \#17 and \#51 (green curve). The identified start, maximum and end time of the impulsive dimming phase are marked by vertical lines. 

This approach for automatically detecting the impulsive phase of the dimming is successful for the majority of events, however for eight events we adjusted either the start or the end manually. Misidentified onsets arise e.g. from local minima in the rising phase of the area growth rate profiles. Misidentifications in the automatic detection of the end of the impulsive phase of the dimming are e.g. due to follow-up events occurring within the time range of the event under study.

In addition to the parameters introduced in Section~\ref{sec:parameters}, we also calculate the duration $t_{\text{dim}}$ of the impulsive phase of the dimming as
\begin{equation}
t_{\text{dim}}=t_{\text{end}}-t_{\text{start}} \;,
\end{equation}
as well as its rise and descend time
\begin{equation}
\begin{split}
t_{\text{rise}}&=t_{\text{max}}-t_{\text{start}} \;, \\
t_{\text{desc}}&=t_{\text{end}}-t_{\text{max}} \;.
\end{split}
\end{equation}

\subsection{Statistics}\label{sec:statistics}
To quantify the uncertainties in the characteristic parameters due to the extraction of dimming pixels at a specific threshold, we apply in addition also intensity thresholds that are 5\% lower and 5\% higher (see the shaded regions in Figures~\ref{fig:profiles_20110621} and~\ref{fig:profiles_20120606}). 
We calculate the timing of the impulsive phase and the characteristic dimming parameters for these representations as well. The central parameter values in the scatter plots shown in the Results section represent the mean, while the error bars reflect the 1$\sigma$ standard deviation.

We calculate the distributions of each characteristic dimming parameter and check whether the variables are log normally distributed. 
The probability function $f(x)$ of the log normal distribution can be written as
\begin{equation}
f(x)=\frac{1}{\sqrt{2\pi}\sigma x}\exp\left(-\frac{(\ln(x)-\mu)^{2}}{2\sigma^{2}}\right) \; ,
\end{equation}
where $\mu$ is the mean and $\sigma$ is the standard deviation of the natural logarithm of $x$ \citep{Limpert:2001,Bein:2011}. 
We use $\mu^{*}=e^{\mu}$ as the median and $\sigma^{*}=e^{\sigma}$ as the multiplicative standard deviation. The confidence interval of 68.3\% is given as \mbox{$[\mu^{*}/\sigma^{*},\mu^{*}\cdot\sigma^{*}]$}.

To investigate how characteristic dimming parameters are statistically related among each other and with basic flare quantities, we calculate the Pearson correlation coefficient in $\log-\log$ space.
We estimate the errors in the correlation coefficient $c$ by using a bootstrapping method: we select $N$\textit{-out-of-}$N$ random data pairs with replacement and calculate $c$, which is repeated for $\sim$ 10000 times and the mean and standard deviation is calculated for the full set as $\bar{c}\pm \Delta c$ \citep{Wall:2012}. 
Following \cite{Kazachenko:2017}, we describe the qualitative strength of the correlation using the following  guide for the absolute value of $c$: $c=[0.2,0.4[$ -- weak, $c=[0.4,0.6[$ -- moderate, $c=[0.6,0.8[$ -- strong and $c=[0.8,1.0]$ -- very strong. 
For parameter combinations where $c>0.4$, we apply a linear regression fit between X and Y in $\log-\log$ space with
\begin{equation}
\log(Y)=k\log(X)+d \; ,
\end{equation}
where $k$ and $d$ represent the coefficients of the regression line, respectively.
\section{Results}
\subsection{Multi-wavelength detections}
We extract coronal dimmings in all seven SDO/AIA EUV filters for all 62 events under study.
For the majority of events, dimming regions are observed in the quiet Sun coronal temperature channels (193 and 211 \AA: 100\%; 171 \AA: 92\%). Although the peak formation temperature of the 335 \AA~channel is $\approx 2.5\times10^{6}$~K, its temperature response curve shows also contributions from cooler temperatures. This might be the reason why the extraction of coronal dimming regions for 94\% of the events was possible. Also in filters sensitive to high temperature plasma (i.e.~94 and 131 \AA) coronal dimmings could be identified for about half of the events (94 \AA: 63\%; 131 \AA: 47\%). Only 15\% of the events show coronal dimmings in wavelengths sensitive to chromospheric plasma (i.e.~in \mbox{304 \AA}). We find that within the analyzed filters, \mbox{211 \AA}~is best suited for the statistical analysis of our event set. Therefore, the calculated parameter values, histograms and correlations presented in the following sections are all based on results using the \mbox{211 \AA}~channel.

\subsection{Core dimming characteristics}\label{sec:results_core}
\begin{figure*}
\centering
\includegraphics[width=1.0\textwidth]{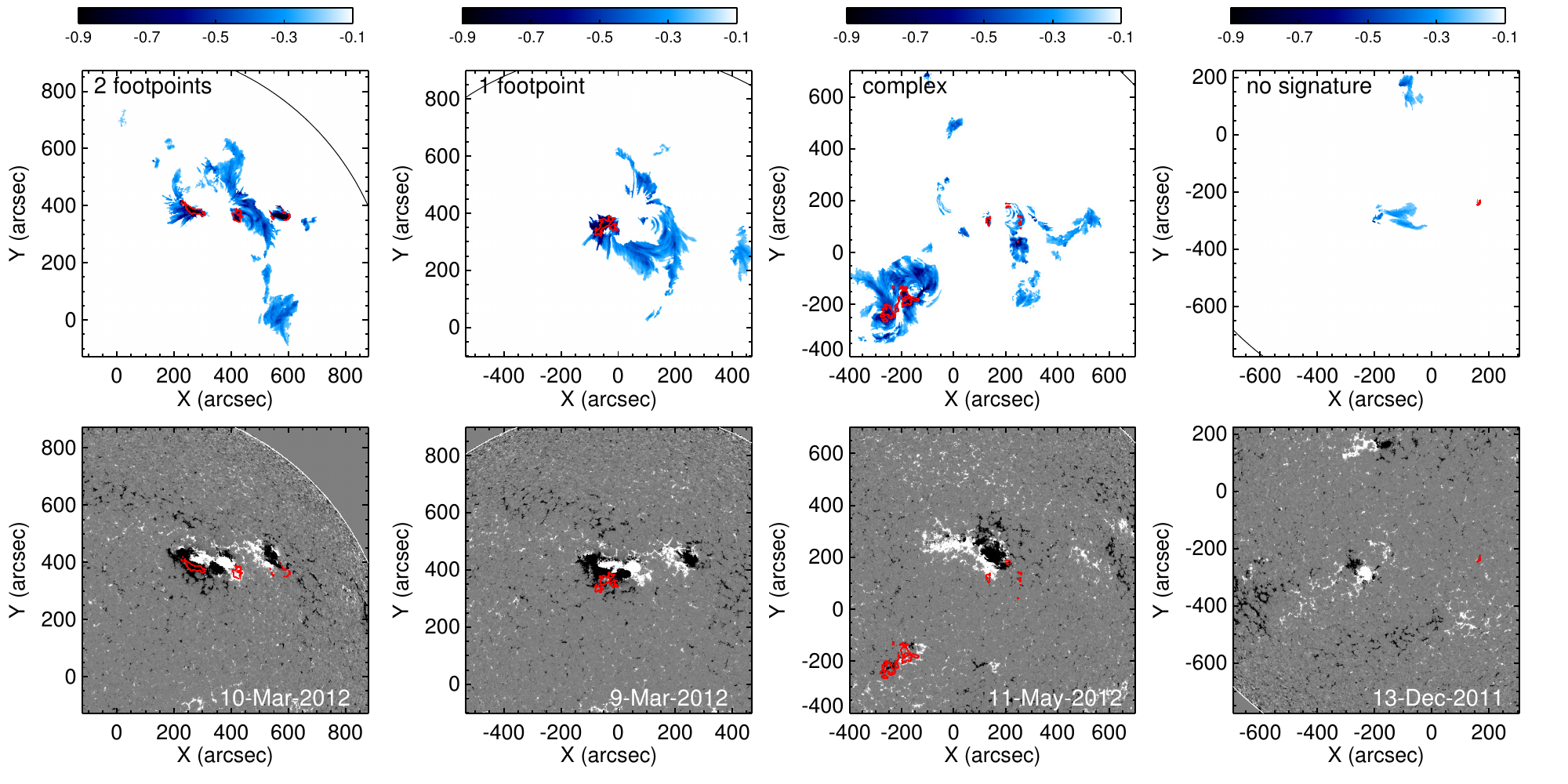}
\caption{Overview of the detection of the four different categories of core dimmings: 2 footpoints, 1 footpoint, complex and no signature, respectively. For each category the logarithmic base-ratio minimum intensity map (top) together with the SDO/HMI LOS magnetogram (bottom) of an example event is plotted. The red contours mark the identified core dimming signatures.}
\label{fig:core_dimmings}
\end{figure*}
We detect potential core dimming regions in SDO/AIA 211 \AA~as described in Section~\ref{sec:detection}. Next to the classical core dimming configuration (i.e.~two footpoint regions) which could be identified for 22 events (35\%), we find three other categories.  For 23\% of the events core dimming signatures are observed in only one predominant magnetic polarity, while 11\% of the events showed more complex signatures, where it was not possible to identify only two isolated regions in opposite magnetic polarities. In 31\% of the cases no core dimming could be identified.
Figure~\ref{fig:core_dimmings} gives an overview of the different categories. The top panels show minimum intensity maps of logarithmic base-ratio data for selected events within our catalog. The red contours mark the detected core dimming regions. In the bottom panels the corresponding SDO/HMI LOS magnetograms together with the contours of the core dimming are shown.
We calculate the area and the total unsigned magnetic flux of the core dimmings only for those events that showed the classical two-footpoint configuration (see~Table~\ref{tab:dimming}).
\subsection{Characteristic dimming parameters and their distributions}
For all 62 coronal dimming events the impulsive phase of the dimming evolution could be uniquely determined from 211 \AA~data and characteristic dimming parameters are calculated.
The minimum in the brightness evolution of the dimming $I_{\text{cu,diff}}$ and its derivative $\dot{I}_{\text{cu,diff}}$ could be identified only for 51 events, due to the occurrence of other events within the time range under study.

In Table~\ref{tab:dimming} we list all parameter values derived. For simplicity only the central values of each parameter are listed.
\renewcommand{\arraystretch}{0.8}
\begin{ThreePartTable}
\begin{TableNotes} \footnotesize
\item \textbf{Note.} For each event we list the dimming area $A$, the maximal area growth rate $\dot{A}$, the total unsigned magnetic flux $\Phi$, the total magnetic flux rate $\dot{\Phi}$, the positive magnetic flux $\Phi_{+}$, the absolute negative magnetic flux $|\Phi_{-}|$, the mean unsigned magnetic flux density $\bar{B}_{\text{us}}$, the total dimming brightness (calculated from minimum intensity maps $I_{\text{min,diff}}$ as well as from the time evolution of $I_{\text{cu,diff}}$), the maximal (negative) brightness change rate $\dot{I}_{\text{cu,diff}}$, the duration of the impulsive phase of the dimming $t_{\text{dim}}$, the core dimming area $A_{\text{core}}$ and its total unsigned magnetic flux $\Phi_{\text{core}}$. Events marked with * are not associated with an EUV wave.
\end{TableNotes}
\begin{longtable*}{p{0.3cm}lp{0.7cm}p{1.1cm}p{0.7cm}p{1.1cm}p{0.7cm}p{0.7cm}p{0.9cm}p{0.9cm}p{0.9cm}p{1.1cm}p{0.7cm}p{0.7cm}p{0.7cm}}
\caption{Results of characteristic dimming parameters} \label{tab:dimming}\\
\toprule
\# & Date &$A$ [km$^{2}$] & $\dot{A}$ [km$^{2}$~s$^{-1}$] & $\Phi$ [Mx]& $\dot{\Phi}$ [Mx~s$^{-1}$]&$\Phi_{+}$ [Mx] &$|\Phi_{-}|$ [Mx] &$\bar{B}_{\text{us}}$ [G] & $I_{\text{min,diff}}$ [DN] & $I_{\text{cu,diff}}$ [DN] & $\dot{I}_{\text{cu,diff}}$ [DN~s$^{-1}$]& $t_{\text{dim}}$ [min]& $A_{\text{core}}$ [km$^{2}$]& $\Phi_{\text{core}}$ [Mx] \\
& &($10^{10}$)&($10^{7}$)&($10^{21}$)&($10^{18}$)&($10^{21}$)&($10^{21}$)& &($10^{5}$)&($10^{5}$)&($10^{3}$)& &($10^{8}$)&($10^{20}$) \\ \midrule

1&20100716&1.16&0.90&0.39&0.32&0.52&0.26&32.19&-5.98&-2.62&-0.17&44.0&-&- \\
2&20100801&9.33&3.48&8.29&4.31&9.79&6.78&57.08&-70.98&-47.98&-2.29&122.3&-&- \\
3&20100807&3.97&4.64&2.37&2.34&3.46&1.29&32.59&-37.57&-25.25&-2.58&29.7&-&- \\
4&20101016&1.30&0.75&0.93&0.91&1.39&0.48&38.28&-10.80&-8.05&-1.70&56.3&-&- \\
5&20101111&0.37&0.41&0.21&0.27&0.23&0.19&52.41&-2.04&-&-&26.7&0.97&0.58 \\
6&20110213&1.99&0.94&2.90&2.23&2.86&2.94&141.92&-14.70&-10.85&-1.54&70.3&6.47&5.81 \\
7&20110214&0.20&0.57&0.52&1.45&0.85&0.19&278.40&-2.77&-&-&11.7&-&- \\
8&20110214&1.06&0.51&1.64&0.86&1.85&1.43&128.39&-10.65&-&-&97.0&4.04&2.20 \\
9&20110214&1.99&1.66&2.97&2.09&3.51&2.44&137.59&-23.60&-&-&54.7&-&- \\
10&20110215&3.60&1.09&3.80&2.07&3.94&3.67&107.77&-21.46&-&-&87.0&10.64&9.69 \\
11&20110215&1.14&1.24&3.68&4.25&4.82&2.54&200.62&-34.58&-&-&24.7&8.48&10.04 \\
12&20110215&1.16&0.82&1.09&1.08&0.63&1.55&91.60&-8.18&-6.98&-0.77&49.0&3.32&0.34 \\
13&20110307&2.94&1.77&1.45&0.77&1.27&1.62&40.13&-20.98&-9.33&-1.56&79.3&-&- \\
14&20110308&0.27&0.27&0.12&0.14&0.13&0.11&48.38&-0.96&-0.59&-0.76&30.0&1.73&0.50 \\
15&20110325&0.76&1.27&0.12&0.24&0.10&0.13&20.80&-2.20&-1.56&-0.46&20.7&-&- \\
16&20110602&3.12&0.77&3.15&1.97&2.86&3.44&75.70&-30.01&-15.98&-1.86&133.3&20.48&7.48\\
17*&20110621&5.24&1.76&3.15&1.25&2.58&3.71&66.12&-55.88&-38.55&-1.14&108.0&48.43&6.62 \\
18&20110702&1.11&1.33&0.93&1.10&1.28&0.58&64.85&-7.22&-3.57&-0.36&38.0&3.89&1.15 \\
19&20110711&4.19&1.79&2.05&0.75&1.49&2.60&52.10&-22.83&-9.61&-0.91&84.3&-&- \\
20&20110802&2.75&0.91&2.17&1.68&1.98&2.35&75.49&-23.61&-10.45&-1.29&86.3&-&- \\
21&20110803&4.20&2.18&4.50&3.15&5.14&3.87&74.30&-42.44&-28.83&-2.31&111.3&-&- \\
22&20110906&5.67&2.68&3.85&3.09&4.32&3.38&68.61&-31.98&-25.44&-3.70&74.3&16.83&7.43 \\
23&20110906&8.45&7.67&7.97&7.28&6.20&9.75&79.12&-71.07&-58.41&-24.60&52.0&29.20&4.12 \\
24&20110908&1.39&0.53&2.94&2.30&1.09&4.79&113.52&-17.00&-&-&85.7&11.01&2.81 \\
25&20110926&1.97&1.31&1.54&0.86&2.16&0.92&61.05&-21.96&-18.52&-2.32&54.3&-&- \\
26&20110927&3.15&2.13&2.50&2.62&4.31&0.68&89.62&-22.41&-17.06&-1.09&53.7&-&- \\
27&20110930&3.55&1.56&1.02&0.54&1.07&0.97&46.11&-19.91&-&-&98.0&-&- \\
28&20111001&5.43&4.08&1.33&0.76&1.76&0.89&37.34&-19.54&-10.95&-2.07&47.3&-&- \\
29&20111002&4.67&4.53&0.87&0.97&1.02&0.72&31.90&-14.17&-10.42&-1.57&40.7&-&- \\
30&20111002&3.27&1.79&0.90&0.46&1.24&0.57&33.36&-10.06&-6.27&-0.38&86.0&-&- \\
31&20111010&0.13&0.26&0.38&0.56&0.11&0.64&195.89&-2.45&-1.86&-0.12&12.3&-&- \\
32&20111114&0.13&0.38&0.02&0.07&0.03&0.01&20.80&-0.56&-0.38&-0.12&11.0&-&- \\
33&20111124&3.05&1.80&2.06&1.06&2.80&1.32&50.60&-29.37&-22.38&-1.40&67.0&-&- \\
34&20111213&0.57&0.38&0.20&0.27&0.16&0.24&43.83&-2.23&-1.63&-0.25&37.0&-&- \\
35&20111222&1.13&0.85&1.33&1.73&0.58&2.07&107.33&-9.07&-5.53&-0.63&41.3&2.87&0.44 \\
36&20111225&2.71&1.91&1.90&1.41&1.97&1.84&64.26&-17.98&-8.97&-1.79&45.7&23.89&7.30 \\
37&20111225&2.16&1.56&2.06&1.99&2.55&1.56&70.12&-25.76&-16.38&-5.28&44.7&23.68&8.83 \\
38&20111225&0.80&0.52&1.39&1.16&1.59&1.20&122.92&-12.61&-8.55&-1.55&50.7&8.50&4.90 \\
39&20111226&1.68&1.84&1.22&1.12&1.29&1.16&51.44&-10.68&-6.33&-0.82&38.7&-&- \\
40&20111226&1.86&1.97&1.70&1.90&1.26&2.13&60.88&-22.57&-12.43&-1.56&47.0&23.35&3.14 \\
41&20120123&4.78&4.04&2.99&2.88&4.13&1.86&40.00&-34.11&-25.96&-4.88&44.3&-&- \\
42&20120306&6.66&4.00&8.31&8.66&12.67&3.95&77.11&-59.09&-&-&48.7&-&- \\
43&20120309&3.30&4.39&3.42&4.12&1.20&5.64&88.70&-26.47&-14.67&-6.02&40.7&-&- \\
44&20120310&4.01&3.26&7.82&5.59&3.68&11.95&107.49&-51.87&-34.09&-3.57&50.7&12.86&19.86 \\
45&20120314&2.83&2.26&2.30&1.88&2.89&1.70&75.43&-30.30&-23.32&-3.29&51.7&-&- \\
46&20120317&0.93&0.89&0.56&0.53&0.77&0.36&56.12&-5.56&-4.44&-1.22&49.0&-&- \\
47&20120405&2.91&3.64&0.87&1.07&0.70&1.04&23.02&-14.82&-12.21&-1.23&31.3&-&- \\
48&20120423&0.34&0.45&0.13&0.24&0.20&0.07&37.82&-5.21&-1.91&-0.54&31.0&-&- \\
49*&20120511&5.28&2.40&3.28&1.73&3.55&3.00&65.16&-41.70&-17.40&-1.05&108.3&-&- \\
50&20120603&3.85&2.46&1.97&1.18&1.50&2.44&35.83&-16.80&-13.60&-9.40&52.0&-&- \\
51&20120606&3.41&2.98&0.82&0.69&0.73&0.90&27.92&-16.14&-10.92&-1.83&47.0&-&- \\
52*&20120614&4.20&1.39&10.82&4.03&9.24&12.40&136.62&-49.79&-31.12&-1.11&131.3&17.13&7.57 \\
53&20120702&4.58&2.23&3.00&3.16&1.95&4.05&63.68&-34.48&-20.31&-5.71&59.3&-&- \\
54&20120702&1.54&1.72&3.18&1.83&2.75&3.61&118.84&-27.68&-15.27&-5.41&42.7&-&- \\
55&20120704&1.28&1.99&0.87&0.98&0.56&1.19&51.14&-17.05&-12.11&-4.28&28.7&-&- \\
56&20120712&3.55&1.41&9.02&3.90&7.52&10.52&121.08&-59.24&-&-&75.3&-&- \\
57&20120813&0.73&1.56&0.11&0.28&0.05&0.17&27.12&-1.52&-1.10&-0.79&16.0&-&- \\
58&20120813&1.10&0.53&0.60&0.52&0.72&0.47&56.20&-5.05&-3.98&-1.04&51.7&-&- \\
59&20120815&1.08&0.76&0.35&0.29&0.32&0.37&26.81&-4.41&-2.42&-0.66&41.7&-&- \\
60*&20120902&4.79&1.15&6.22&1.73&6.51&5.94&104.97&-56.32&-41.63&-1.62&176.0&25.04&14.15 \\
61&20120925&1.29&1.88&0.63&0.77&0.40&0.87&48.48&-15.92&-&-&32.7&10.02&0.92 \\
62&20120927&1.62&1.04&2.50&1.42&3.49&1.52&88.48&-23.56&-8.49&-1.03&77.7&-&- \\
\bottomrule
\insertTableNotes\\
\end{longtable*}
\end{ThreePartTable}

\renewcommand{\arraystretch}{1.25}
\begin{table*}
\centering
\begin{tabular}{p{3.2cm} l l l l l}
\toprule
   \multicolumn{2}{c}{parameter} & typical range & mean $\pm$ STD & $\mu^{*}$ & confidence interval\\ \hline
  area & $A$ [km$^{2}$] & $[0.13,9.33]\;\times10^{10}$ & $2.70\pm2.00 \times 10^{10}$ & $2.15\times10^{10}$ & $[0.82, 5.61]\;\times 10^{10}$\\
   area growth rate & $\dot{A}$ [km$^{2}$~s$^{-1}$] & $[0.26,7.67]\;\times10^{7}$ & $1.83\pm1.38 \times 10^{7}$ & $1.38\times10^{7}$ & $[0.62, 3.11]\;\times 10^{7}$\\ 
     magnetic area & $A_{\phi}$ [km$^{2}$] & $[0.20,29.0]\;\times10^{9}$ & $6.67\pm5.66 \times 10^{9}$ & $5.35 \times 10^{9}$ & $[1.75,16.4]\;\times 10^{9}$\\
     magnetic area growth rate & $\dot{A}_{\phi}$ [km$^{2}$s$^{-1}$] & $[0.34,20.3]\;\times10^{6}$ & $4.41\pm4.07 \times 10^{6}$ & $5.14 \times 10^{6}$ & $[1.46,7.08]\;\times10^{6}$\\ \hline
   total unsigned magnetic flux & $\Phi$ [Mx] & $[0.02,10.8]\;\times10^{21}$ & $2.442\pm2.443 \times 10^{21}$ & $1.75 \times 10^{21}$ & $[0.58, 5.30]\; \times 10^{21}$ \\
   total unsigned magnetic flux rate & $\dot{\Phi}$ [Mx~s$^{-1}$]& $[0.67,86.56]\;\times10^{17}$ & $1.79\pm1.67 \times 10^{18}$ & $1.35 \times 10^{18}$ & $[0.46,3.96]\; \times 10^{18}$ \\
      mean unsigned magnetic flux density & $\bar{B}_{\text{us}}$ [G] & $[20.80,278.40]$ & $74.90\pm47.62$ & $60.95$ &$[33.12,112.17]$ \\ \hline
  positive magnetic flux & $\Phi_{+}$ [Mx] & $[0.03,12.7]\;\times10^{21}$ & $2.45\pm2.54 \times 10^{21}$ & $1.66 \times 10^{21}$ & $[0.59,4.65]\;\times 10^{21}$ \\
   negative magnetic flux & $\Phi_{-}$ [Mx] & $[-12.4,-0.01]\;\times10^{21}$ & $-2.44\pm2.79 \times 10^{21}$ & $-1.58\times 10^{21}$ & $[-4.84,-0.52]\;\times 10^{21}$\\ \hline
   total brightness & $I_{\text{min,diff}}$ [DN] & $[-71.1,-0.56] \times 10^{5}$ & $-2.27\pm1.79 \times 10^{6}$ & $-1.91 \times 10^{6}$ & $[-5.52,-0.66]\;\times 10^{6}$ \\
      (cumulative) total brightness & $I_{\text{cu,diff}}$ [DN] & $[-58.4,-0.38] \times 10^{5}$ & $-1.46\pm1.28 \times 10^{6}$ & $-1.09 \times 10^{6}$ & $[-2.66,-0.45]\;\times 10^{6}$ \\
       brightness change rate & $\dot{I}_{\text{cu,diff}}$ [DN~s$^{-1}$] & $[-24.6,-0.12] \times 10^{3}$ & $-2.42\pm3.67 \times 10^{3}$ & $-1.43 \times 10^{3}$ & $[-3.00,-0.68]\;\times 10^{3}$ \\ \hline
   duration of the impulsive phase & $t_{\text{dim}}$ [min] & $[11.0,176.0]$ & $58.7\pm32.9$ & $50.4$ & $[32.1, 79.0]$\\
  rise time & $t_{\text{rise}}$ [min] & $[6.7,96.3]$ & $20.4\pm16.3$ & $14.2$ &$[8.5,23.6]$ \\ 
  descend time & $t_{\text{desc}}$ [min] & $[2.0,110.3]$ & $38.3\pm24.9$ & $33.5$ &$[19.3, 58.0]$ \\ \hline
  core dimming area (in \% of $A$) & $A_{\text{core}}/A$ [\%] & $[2.6,12.6]$ & $5.9\pm3.1$ & $5.0$ & $[2.6, 9.7]$\\
  core total unsigned magnetic flux (in \% of $\Phi$) & $\Phi_{\text{core}}/\Phi$ [\%] & $[3.1,43.0]$ & $20.8\pm11.8$ & $20.1$ & $[10.2, 39.7]$\\\hline
    mean brightness decrease total dimming & $I_{\text{min,total}}/I_{\text{pre}}$ [\%] & $[48.2,69.7]$ & $58.0\pm5.5$ & -- & --\\
    mean brightness decrease core dimming & $I_{\text{min,core}}/I_{\text{pre}}$ [\%] & $[61.4,81.5]$ & $75.0\pm5.2$ & -- & --\\\hline\hline
\end{tabular}
\caption{Statistical coronal dimming parameters. The typical range for each quantity is given, as well as the mean, the standard deviation (STD), the median $\mu^{*}$ and the 68.3\% confidence interval derived from the log normal fit to the distribution.}
\label{tab:histograms}
\end{table*}

Table~\ref{tab:histograms} gives an overview of statistical parameters calculated from the distributions of all dimming quantities derived: we list the typical ranges, the arithmetic mean with the standard deviation, as well as the median and the 68.3\% confidence interval calculated from the fit parameters of the log normal fit ($\mu$ and $\sigma$) to the histograms.

\begin{figure}
\centering
\includegraphics[width=1.0\columnwidth]{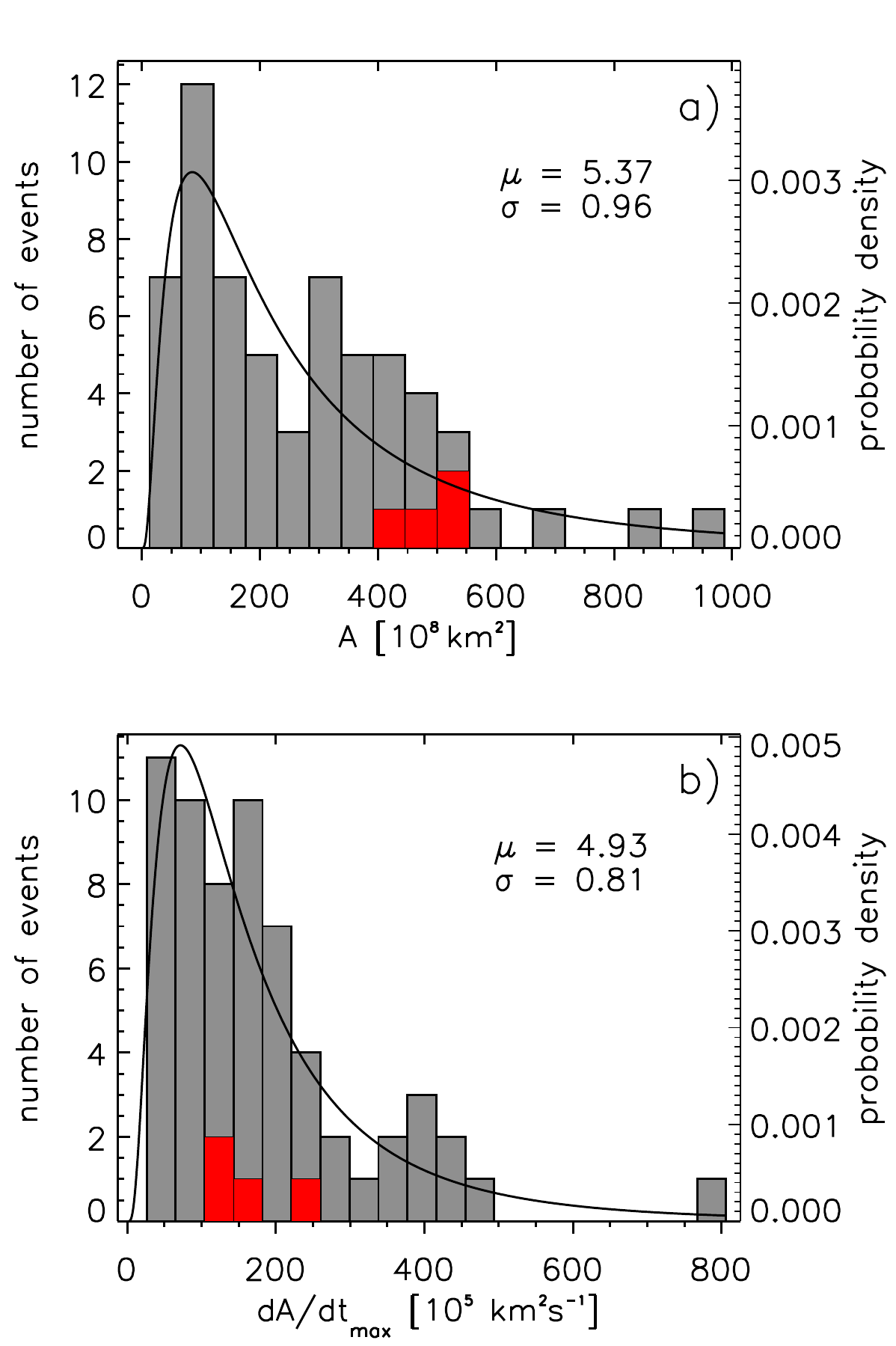}
\caption{Distribution of (a) the dimming area $A$ and (b) the maximal area growth rate $\dot{A}$. The solid lines represent the log normal fit to each distribution. The histogram of dimming events that were not associated with an EUV wave are overplotted in red.}
\label{fig:hist_area}
\end{figure}
\begin{figure}
\centering
\includegraphics[width=1.0\columnwidth]{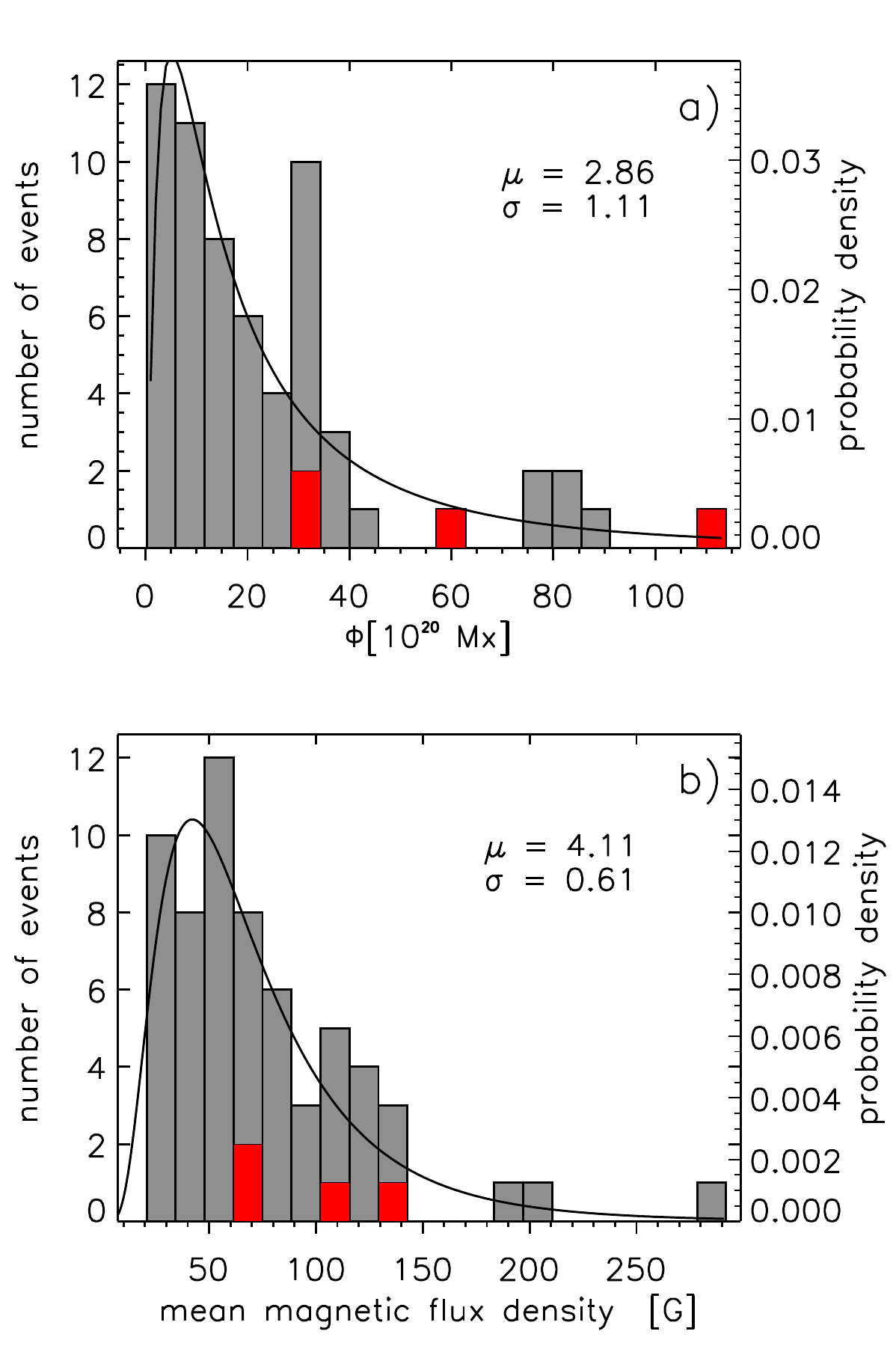}
\caption{Distribution of (a) the total unsigned magnetic flux $\Phi$ of the dimming regions and (b) the mean unsigned magnetic flux density $\bar{B}_{\text{us}}$.}
\label{fig:hist_magnetic_flux}
\end{figure}
\begin{figure}
\centering
\includegraphics[width=1.0\columnwidth]{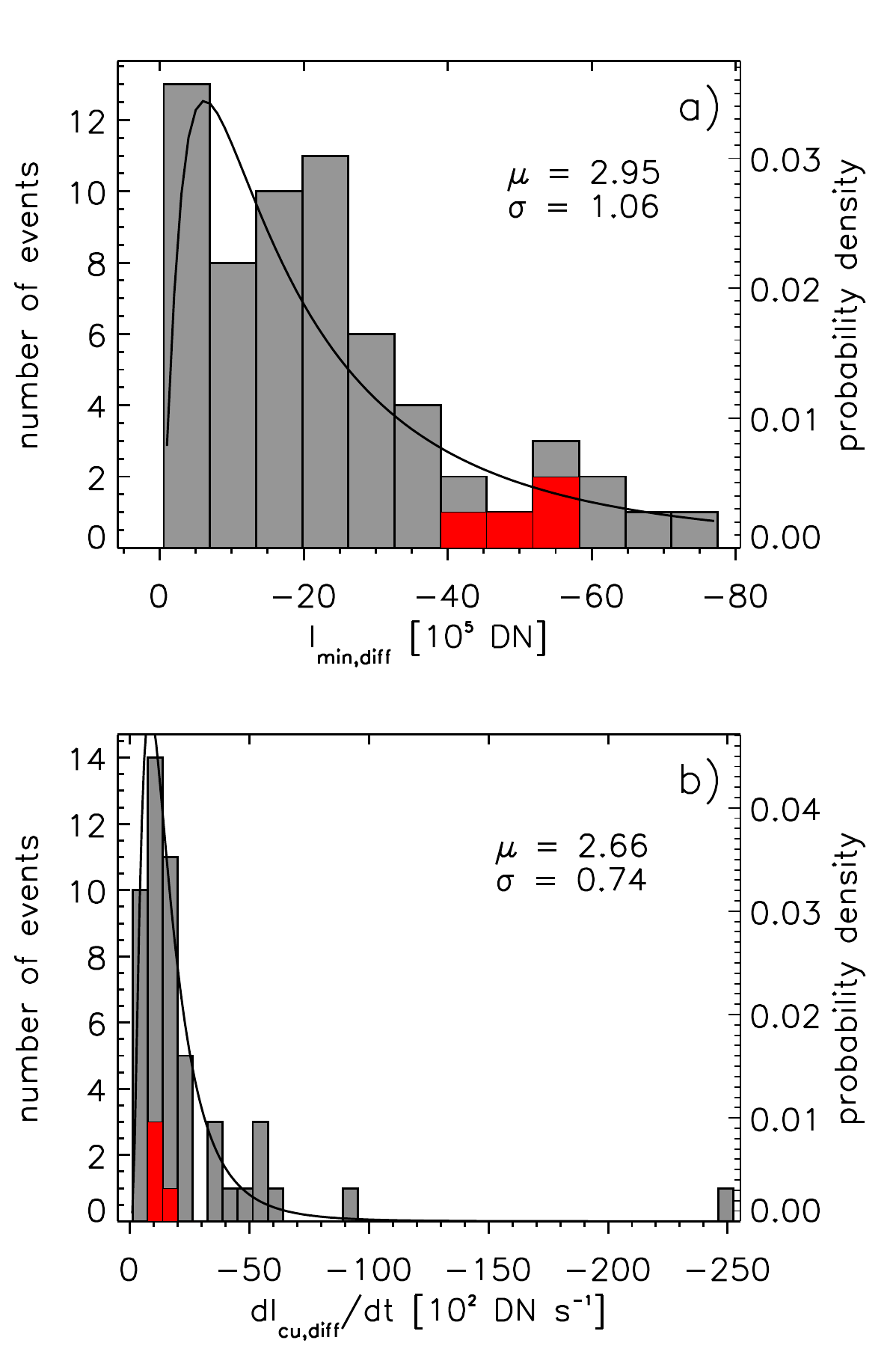}
\caption{Distribution of (a) the total minimum brightness and (b) the maximal brightness change rates.}
\label{fig:hist_brightness}
\end{figure}
\begin{figure}
\includegraphics[width=1.0\columnwidth]{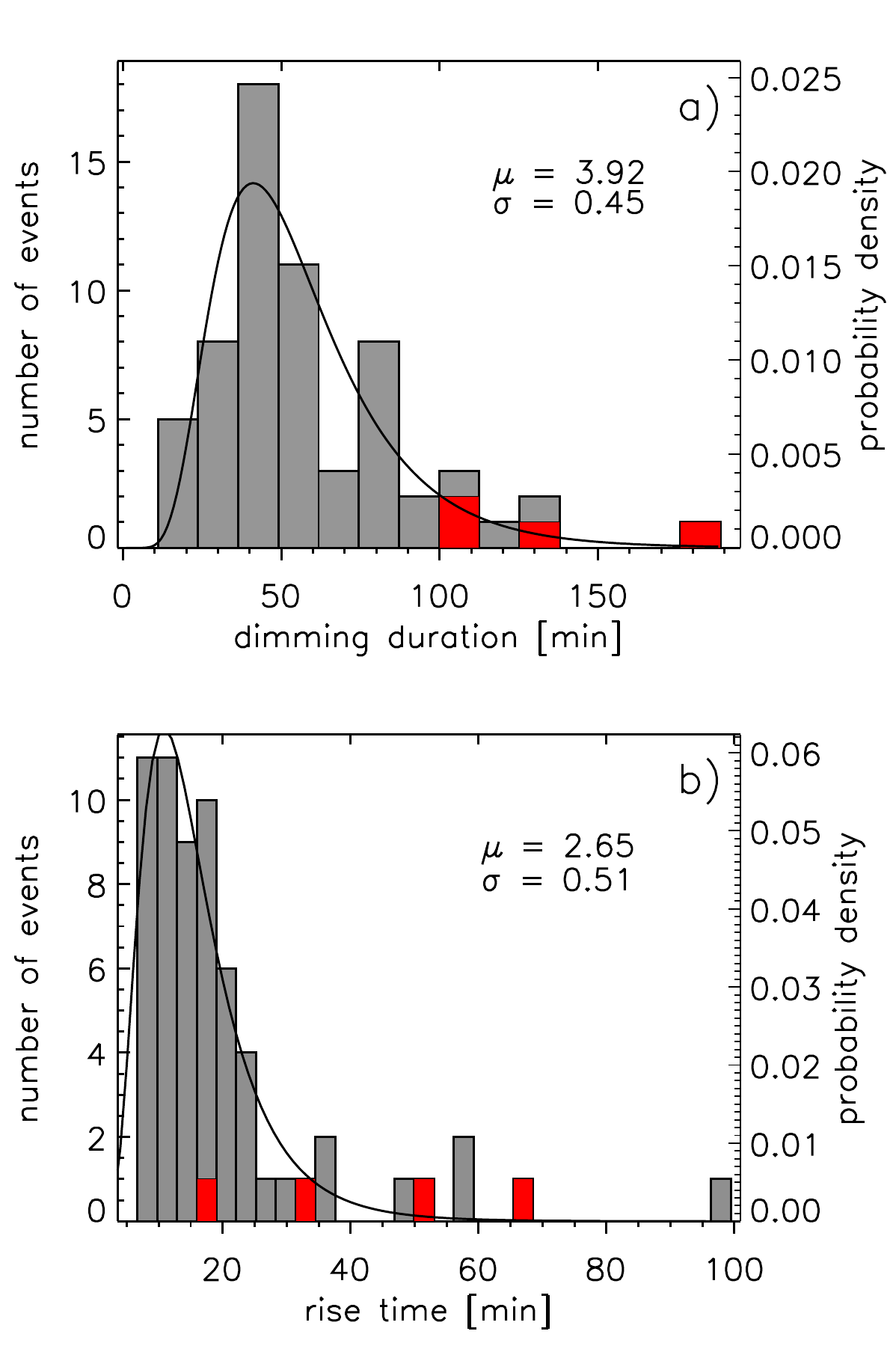}
\caption{Distribution of (a) the duration of the impulsive phase and (b) the rise time of the dimming.}
\label{fig:hist_duration}
\end{figure}
\begin{figure}
\includegraphics[width=1.0\columnwidth]{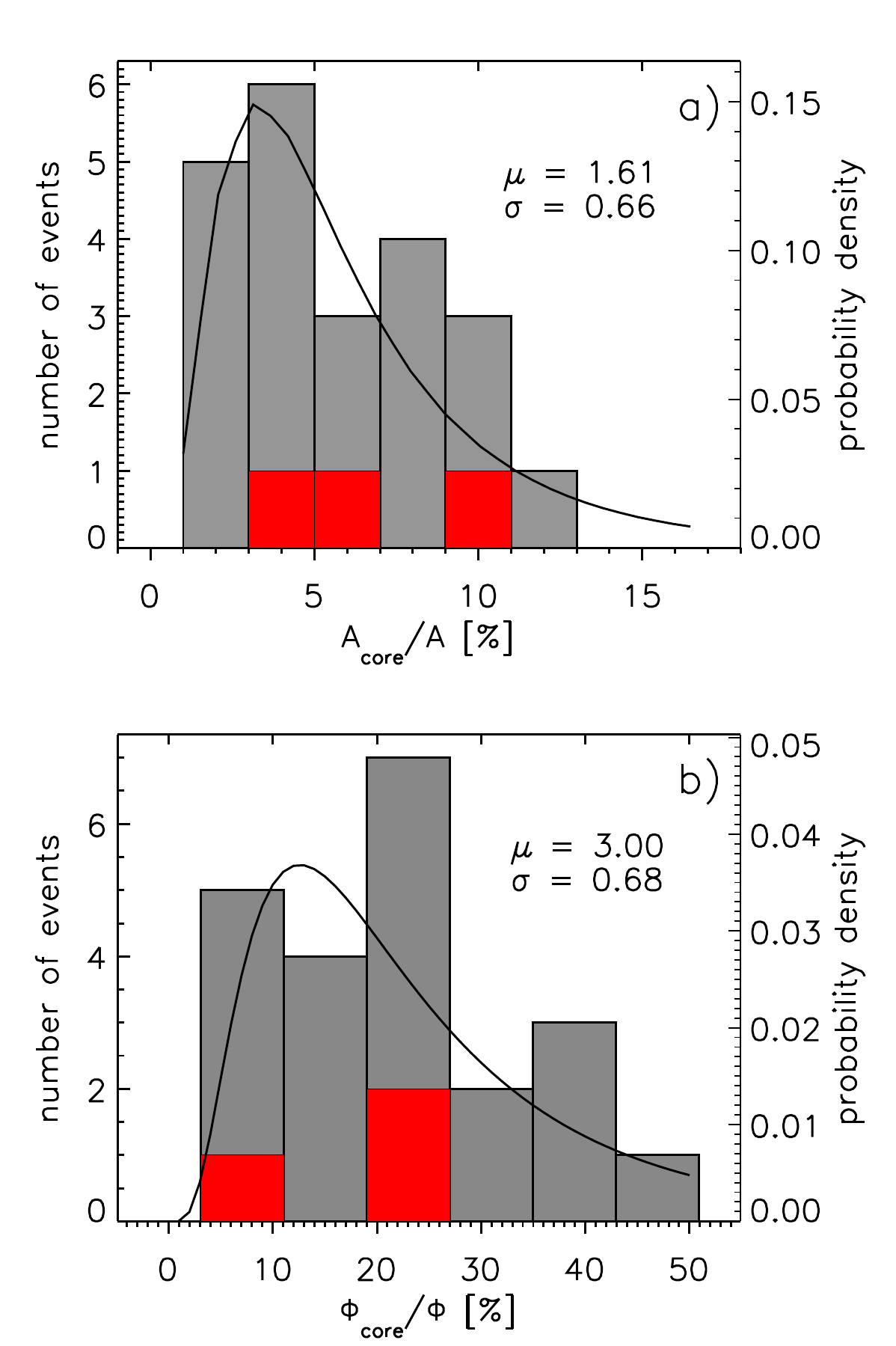}
\caption{Distribution of (a) the core dimming area $A_{\text{core}}$ in \% of the total dimming area $A$ and (b) its total unsigned magnetic flux $\Phi$ in \% of the total magnetic flux.}
\label{fig:hist_core}
\end{figure}
\begin{figure}
\includegraphics[width=1.0\columnwidth]{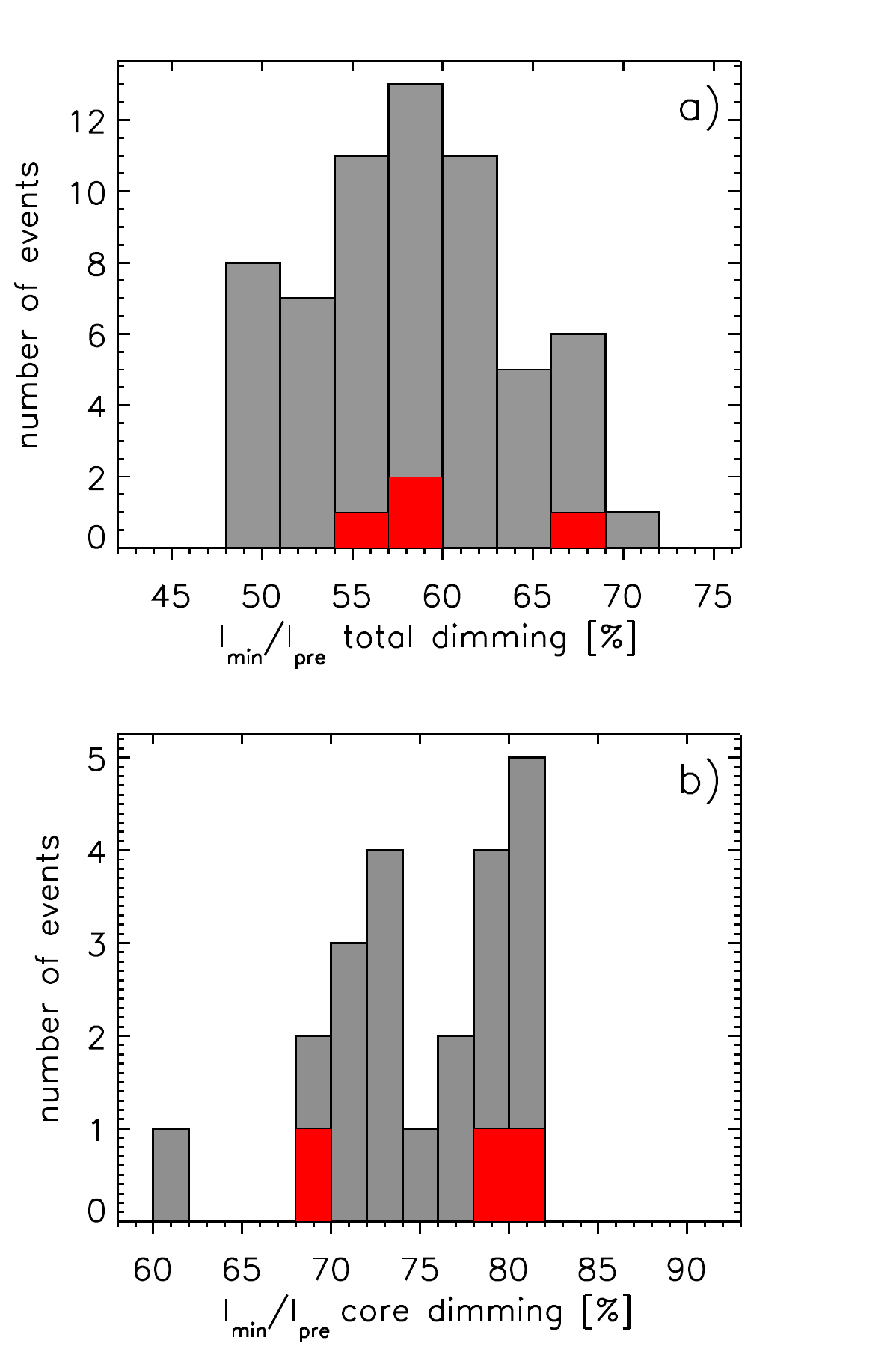}
\caption{Distribution of (a) the mean brightness of the total dimming region and (b) the mean brightness of the core dimming region in \% of the mean pre-event intensity level.}
\label{fig:hist_bright_total_core}
\end{figure}

Figures~\ref{fig:hist_area} -- \ref{fig:hist_bright_total_core} show the distributions of selected dimming parameters for the whole sample (gray histograms) together with the distribution for events not associated with an EUV wave colored in red. All histograms are asymmetrical with a tail towards high values. The log normal probability density function is derived for each distribution and overplotted as solid black line. The fit parameters $\mu$ and $\sigma$ are also given in each plot.

Figure~\ref{fig:hist_area} (a) presents the distribution of the dimming area $A$ for all dimming events under study. The values cover a range over about two orders of magnitude, from $1.30\times 10^{9}$~km$^{2}$ to $9.33\times10^{10}$~km$^{2}$, which is comparable to the typical values found by \cite{Aschwanden:2016}. The majority of events show values $<6.0\times 10^{10}$~km$^{2}$. Only three events reach an area larger than that. The median calculated from the log normal fit to the data is $2.15\times10^{10}$~km$^{2}$ and the confidence interval results in $[0.82,5.61]\times 10^{10}$~km$^{2}$.  
Events that are not associated with an EUV wave (overlaid in red) are located at the right end of the distribution, i.e.~show higher values for the dimming area.

The distribution of the maximal area growth rate $\dot{A}$ is shown in Figure~\ref{fig:hist_area} (b). $\dot{A}$ ranges from \mbox{$2.60\times10^{6}$~km$^{2}$~s$^{-1}$} to $7.67\times10^{7}$~km$^{2}$~s$^{-1}$, with a mean of $1.83\times10^{7}$~km$^{2}$~s$^{-1}$. The median derived from the log normal fit is \mbox{$1.38\times10^{7}$~km$^{2}$~s$^{-1}$} and only one event exceeds $5\times 10^{7}$~km$^{2}$~s$^{-1}$ (\#23, 2011 September 6). Events that are not associated with an EUV wave (indicated in red) do not show any significant difference from the original distribution.

Figure~\ref{fig:hist_magnetic_flux} shows the histograms of the magnetic properties of coronal dimmings, the total unsigned magnetic flux $\Phi$ (panel (a)) and the mean unsigned magnetic flux density $\bar{B}_{\text{us}}$ (panel (b)). More than 90\% of the dimmings have a total unsigned magnetic flux $<5.0\times 10^{21}$~Mx and on average $2.442\pm2.443\times10^{21}$~Mx is involved. The log normal median results in $1.75\times10^{21}$~Mx. As for the dimming area, events that show no signature of an EUV wave are located at the far right side of the histogram.
Panel~(b) shows the distribution of the mean unsigned magnetic flux density $\bar{B}_{\text{us}}$. For 95\% of the events $\bar{B}_{\text{us}}$ is $<$150~G. The mean of the distribution lies at $74.90\pm47.62$~G and the median obtained from the log normal fit is $60.95$~G, with a confidence interval of $[33.12,112.17]$~G. 

Figure~\ref{fig:hist_brightness} shows the distributions of the brightness properties of the coronal dimming events. In panel (a) the total brightness of the dimmings $I_{\text{min,diff}}$, calculated from minimum intensity maps,  is given. Values range over three orders of magnitude from $[-71.1, -0.56]\times 10^{5}$~DN, with a mean at $-2.27\pm1.79 \times 10^{6}$~DN. Events not associated with an EUV wave reach values beyond $-4.0\times10^{6}$~DN.
Figure~\ref{fig:hist_brightness}~(b) shows the histogram of the brightness change rate $\dot{I}_{\text{cu,diff}}$. The distribution shows a well-defined peak at $-13.8-7.5\times 10^{2}$~DN~s$^{-1}$ and 80\% of the events show values beyond $-3.0\times10^{3}$~DN~s$^{-1}$. The mean and median values are $-2.42\pm3.67 \times 10^{3}$~DN~s$^{-1}$ and $-1.43\times10^{3}$~DN~s$^{-1}$, respectively. No differences are found within the distributions for events with/without an EUV wave.

Figure~\ref{fig:hist_duration} presents the distributions of (a) the duration of the impulsive phase of the dimming $t_{\text{dim}}$ and (b) its rise time $t_{\text{rise}}$. $t_{\text{dim}}$ ranges from $11-176$~min, with a mean value of $58.7\pm32.9$~min and a log normal median value of $50.4$~min around a confidence interval of $[32.1,79.0]$~min.
High dimming duration values ($>$100~min), indicative of a gradual evolution, result mainly from events that were not associated with an EUV wave (indicated in red). 
For the rise time $t_{\text{rise}}$, the majority of events (85\%) have values $<$30~min. The mean of the distribution is $20.4\pm16.3$~min and the log normal median is $14.2$~min at a confidence interval of $[8.5,23.6]$~min. Compared to the descend time of the impulsive phase $t_{\text{desc}}$ (see Table~\ref{tab:histograms}) the rise time is much shorter.

Figure~\ref{fig:hist_core} shows the distributions of the core dimming area (panel (a)) and its total unsigned magnetic flux (panel (b)).
On average, core dimming regions contain 20\% of the total unsigned magnetic flux of the total dimming region but covers only 5\% of its total area. For six events, $\Phi_{\text{core}}$ results in $>$30\% of $\Phi$. The mean total unsigned magnetic flux density ranges between $[68.0, 336.03]$~G and its average is $162.65\pm89.03$~G, more than a factor of 2 stronger as for the total dimming region.

Figure~\ref{fig:hist_bright_total_core}~(a) shows the distribution of the mean brightness decrease of the total dimming region with respect to its intensity level before the eruption (in \%). The brightness drops range from $48.2-69.7\%$, the mean is $58.0\pm5.5\%$ and the median $57.5\pm4.4\%$. The distribution of the mean brightness of the core dimming regions is plotted in panel (b). These regions show on average a stronger decrease in their mean intensity by $75.0\pm5.2\%$, with a median of $76.2\pm4.2\%$. Typical values of this distribution range from $61.4-81.5\%$.
\subsection{Correlations of characteristic dimming parameters}
In Figures~\ref{fig:corr_magnetic_flux}--\ref{fig:corr_brightness_mag_flux} the most significant dependencies among the dimming parameters are presented. In each plot, the red crosses correspond to dimming events that were not associated with an EUV wave. The blue lines represent the linear fit to the total distribution in $\log-\log$ space, and the corresponding correlation coefficients are given in the top-left corner of each panel. In addition, Figure~\ref{fig:correlations} lists the correlation coefficients for all possible parameter pairs together with their p-values in matrix form. Note, that not all matrix entries represent physically meaningful or causal relationships.

Figure~\ref{fig:corr_magnetic_flux} shows the total unsigned magnetic flux $\Phi$ as a function of (a) the magnetic area $A_{\Phi}$ and (b) the mean unsigned magnetic flux density $\bar{B}_{\text{us}}$. We find a higher correlation ($c=0.89\pm 0.03$) for the area than for $\bar{B}_{\text{us}}$ ($c=0.53\pm 0.1$). This indicates that the differences in $\Phi$ for different events, which ranges over almost three orders of magnitude, mostly result from differences in the dimming area and less from differences in the underlying magnetic flux densities.
\begin{figure}
\centering
\includegraphics[width=1.0\columnwidth]{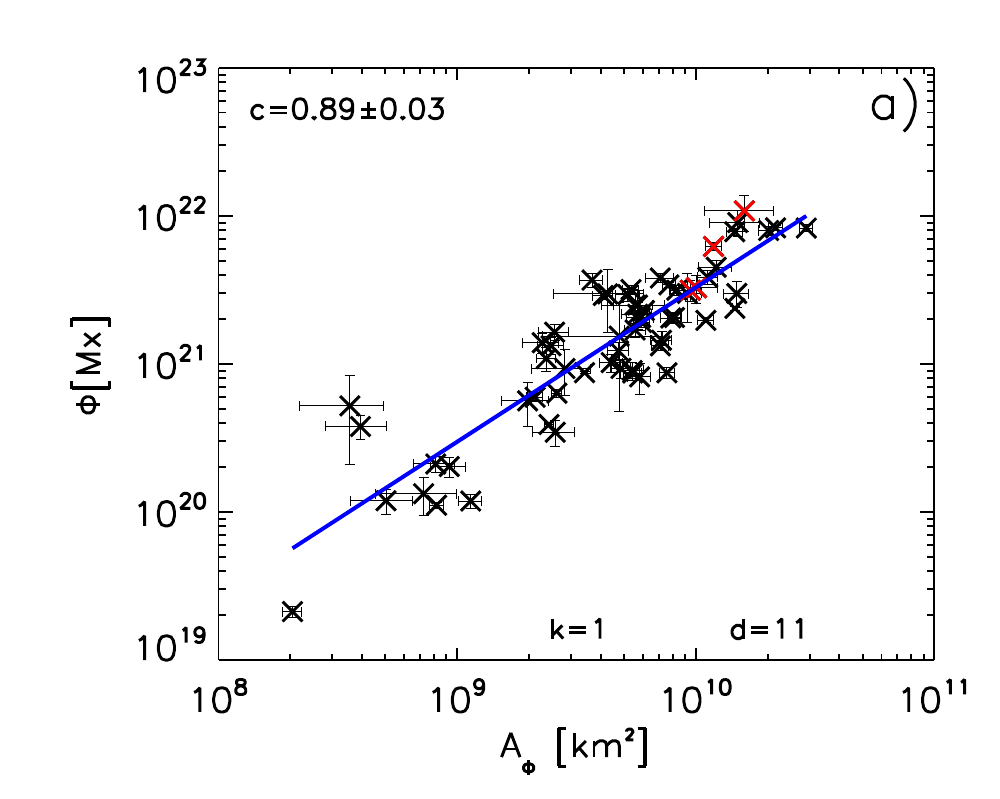}
\includegraphics[width=1.0\columnwidth]{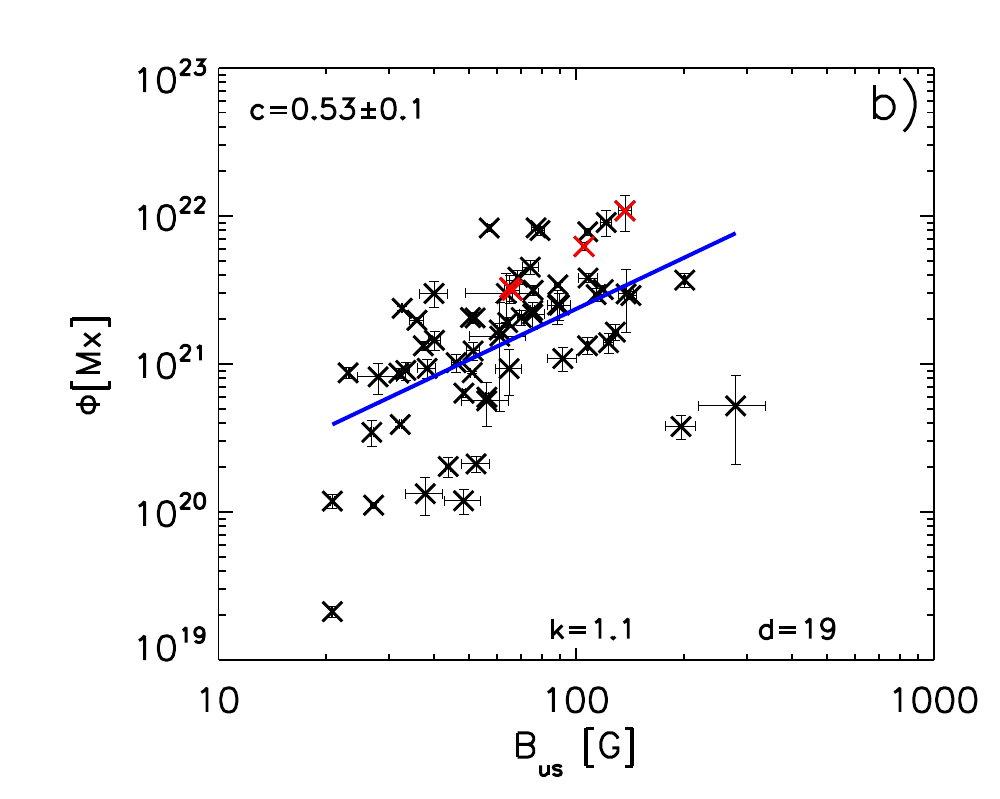}
\caption{Correlation plots showing the total unsigned magnetic flux $\Phi$ against (a) the magnetic dimming area $A_{\phi}$ and (b) the mean unsigned magnetic field density $\bar{B}_{\text{us}}$.}
\end{figure}
\label{fig:corr_magnetic_flux}

\begin{figure}
\centering
\includegraphics[width=1.0\columnwidth]{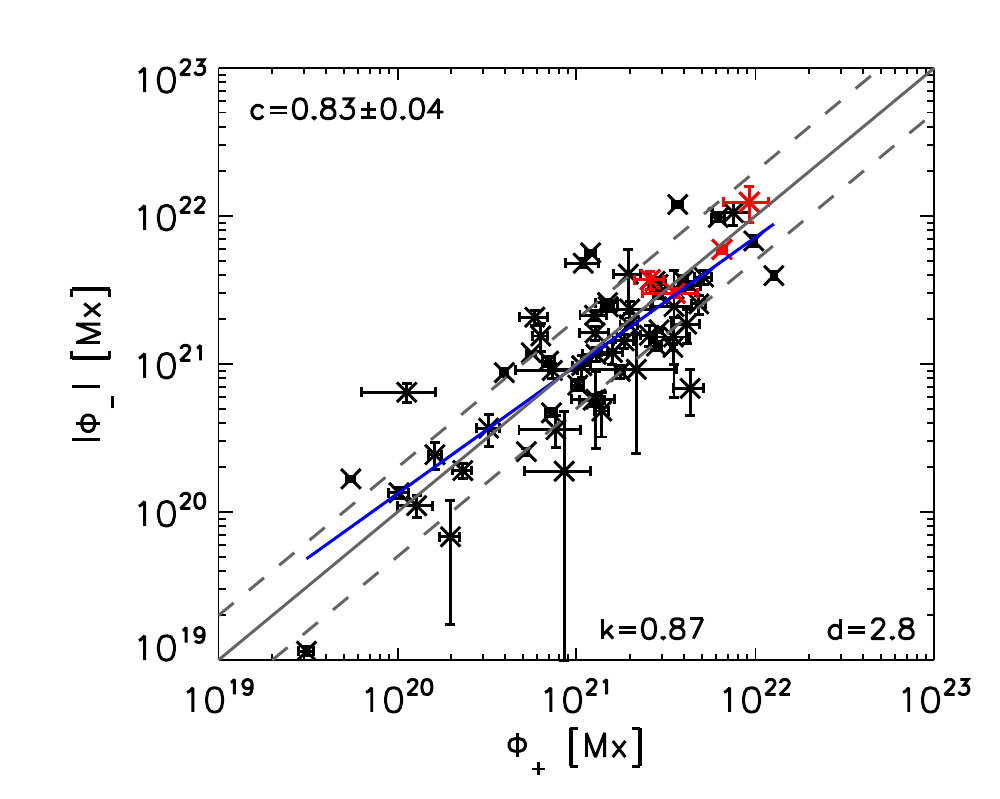}
\caption{Absolute negative magnetic flux $|\Phi_{-}|$ against the positive magnetic flux $\Phi_{+}$ of the coronal dimming regions, together with the regression line (blue line) and the 1:1 correspondence (gray line). The dashed lines indicate ratios of 0.5 and 2 between the fluxes.}
\label{fig:magnetic_flux_balance}
\end{figure}

Figure~\ref{fig:magnetic_flux_balance} shows the scatter plot of the absolute negative magnetic flux $|\Phi_{-}|$ against the positive magnetic flux $\Phi_{+}$, revealing a high correlation of $c=0.83\pm 0.04$. The fitted regression line (blue line) shows a slope ($k$=0.87) close to the 1:1 correspondence (gray line), indicating that the total unsigned magnetic flux $\Phi$ of the dimming regions is balanced. For 65\% of the events the ratio between the positive and negative magnetic flux lies between 0.5 and 2.0 (indicated by the dashed lines). \cite{Qiu:2005} defined this range to be balanced (for total reconnection fluxes obtained from flare observations), since a perfect balance is never obtained due to measurement uncertainties.

\begin{figure}
\centering
\includegraphics[width=1.0\columnwidth]{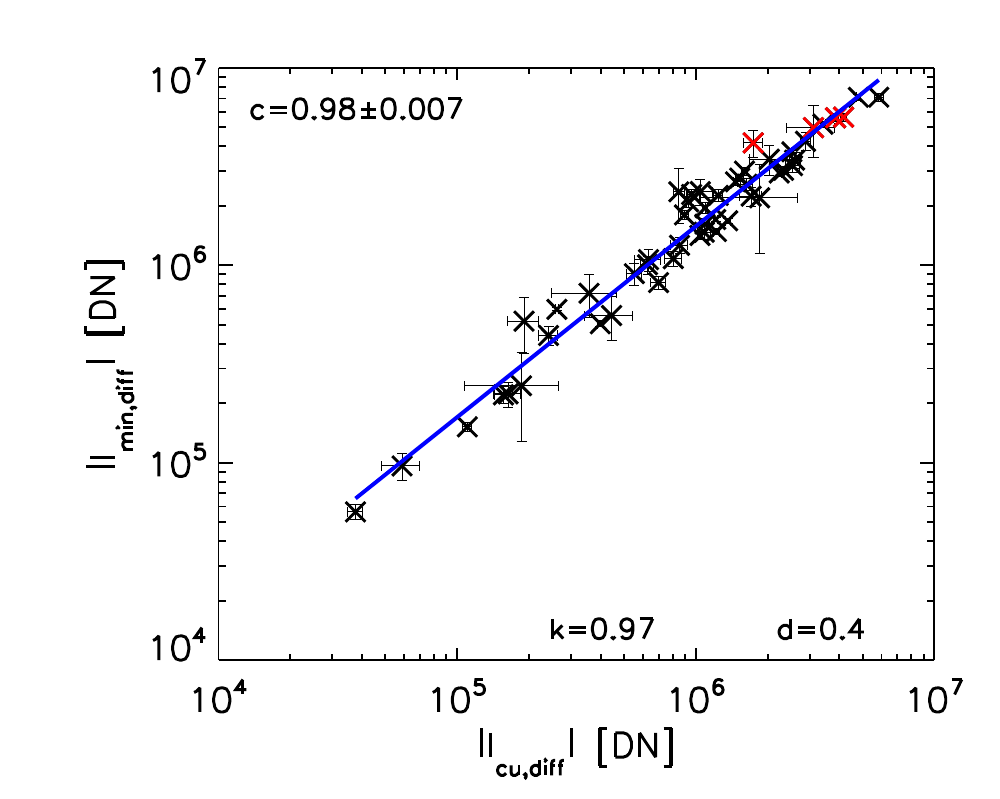}
\caption{Absolute total brightness of the dimmings $|I_{\text{min,diff}}|$ versus the absolute minimum brightness extracted from the time evolution of $I_{\text{cu,diff}}$.}
\label{fig:corr_dimming_brightness}
\end{figure}

\begin{figure}
\centering
\includegraphics[width=1.0\columnwidth]{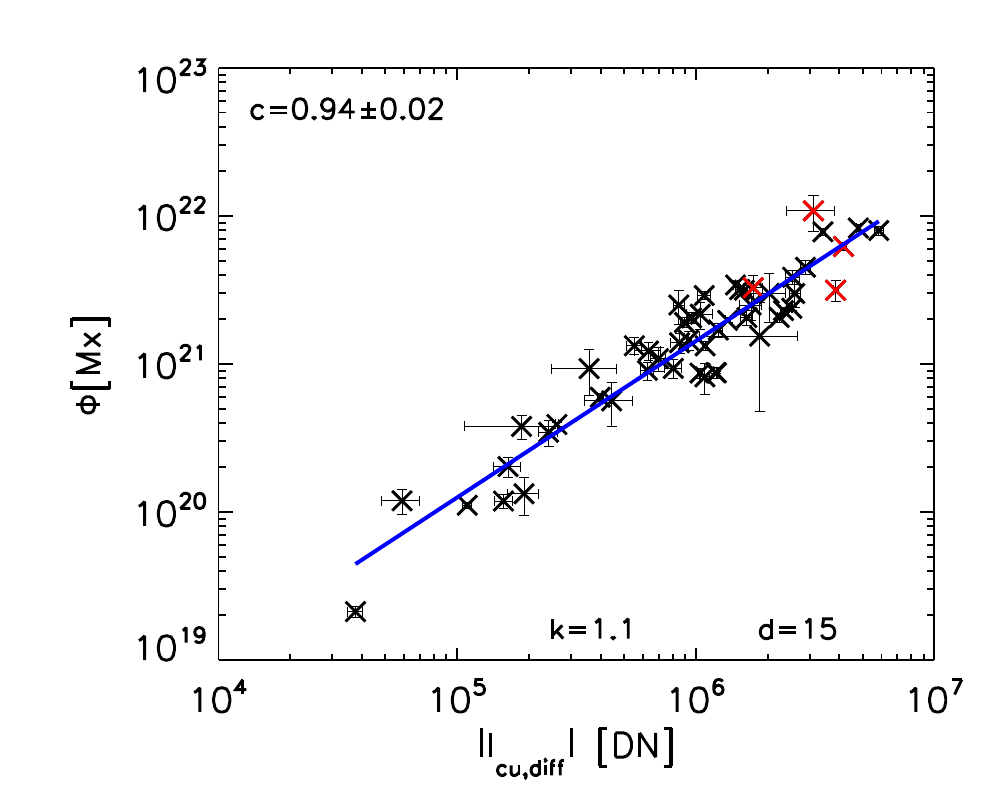}
\caption{Correlation plot between the total unsigned magnetic flux $\Phi$ and the absolute total minimum brightness $|I_{\text{cu,diff}}|$.}
\label{fig:corr_brightness_mag_flux}
\end{figure}

\begin{figure*}
\centering
\includegraphics[width=1.0\textwidth]{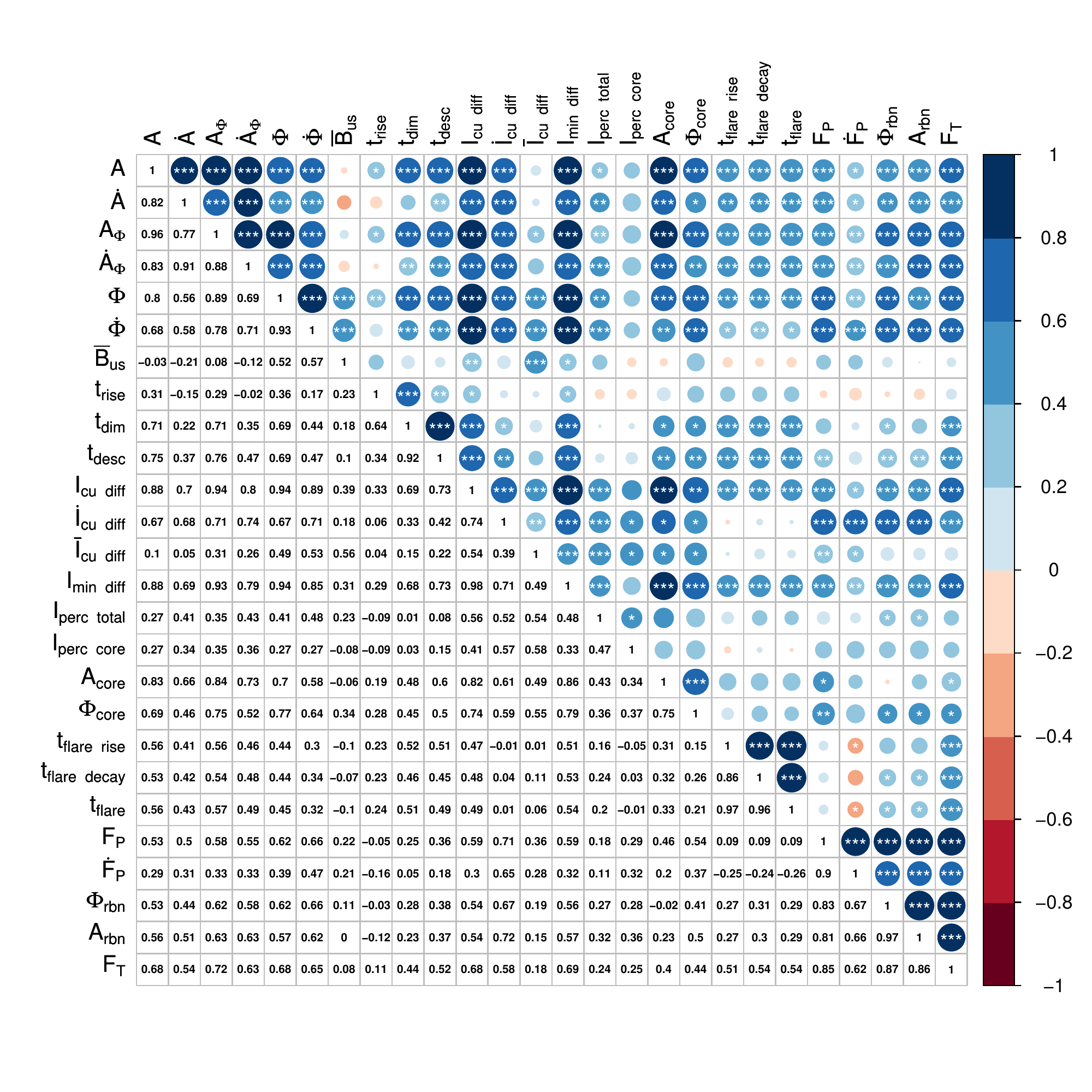}
\caption{Correlation matrix showing the correlation coefficients of all possible pairs of dimming and flare parameters under study. All entries above the main diagonal show a graphical interpretation of the correlation coefficient. The color and the size of the circles indicate the value of the correlation coefficient and if it is positive (blue) or negative (red). The white stars in the circles represent the calculated p-values for each coefficient, where * indicates p-values smaller than 0.05, ** $p\le0.01$ and for *** $p\le 0.001$, respectively. All entries below the main diagonal present values of the correlation coefficients. As the diagonal represents the correlation of each variable with itself it is 1 by definition.}
\label{fig:correlations}
\end{figure*}

Figure~\ref{fig:corr_dimming_brightness} compares the absolute total brightness of the dimmings, on the one hand defined as absolute sum of all pixels within minimum intensity maps $|I_{\text{min,diff}}|$ and on the other hand extracted as absolute minimum brightness from the time evolution of $I_{\text{cu,diff}}$. The highly positive correlation coefficient of $c=0.98\pm 0.007$ and the slope of the fitted regression line close to 1 (k=0.97), reveals that similar values for the dimming brightness are obtained using two different approaches. 

Figure~\ref{fig:corr_brightness_mag_flux} presents the relationship between the total unsigned magnetic flux $\Phi$ and the absolute total dimming brightness $|I_{\text{cu,diff}}|$, showing a very strong correlation of $c=0.94\pm0.02$. We note that both quantities depend on the dimming area, so from this relationship it is not yet clear whether the darkening of the dimming observed in base-difference images results from the strength of the underlying magnetic field or from the size of the dimming region.
Investigating the relationship between their area-independent quantities, reveals a moderate positive correlation between the mean unsigned magnetic flux density $\bar{B}_{\text{us}}$ and the absolute mean brightness $|\bar{I}_{\text{cu,diff}}|$, representing the mean intensity of each dimming pixel ($c=0.56\pm0.09$). The absolute mean brightness however does not correlate with the size of the dimming region ($c=0.1\pm0.2$). This indicates that the stronger the underlying magnetic field, the darker the dimming. 
We also note that the percentage drop in the mean brightness compared to its pre-eruption intensity level does not show any significant correlation with any other dimming parameter.

\subsection{Relations between dimming and flare parameters}
In order to identify general relationships with the associated flares we correlate various characteristic dimming parameters with the GOES SXR peak flux $F_{P}$, the maximum of its derivative $\dot{F}_{P}$, its SXR fluence $F_{T}$, and the duration of the flare $t_{\text{flare}}$. For event \#32 only a small increase in the SXR flux in the descending phase of a larger flare could be identified and no flare parameters could be obtained. Therefore, this event was excluded from this part of the statistics.

\begin{figure}
\centering
\includegraphics[width=1.0\columnwidth]{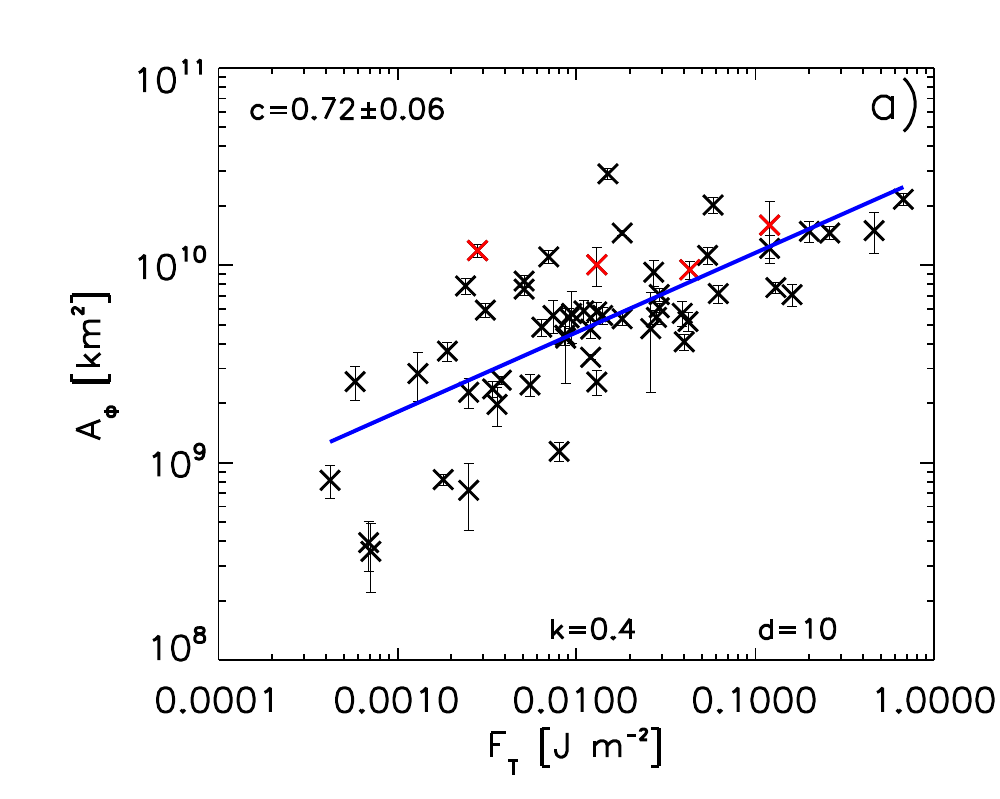}
\includegraphics[width=1.0\columnwidth]{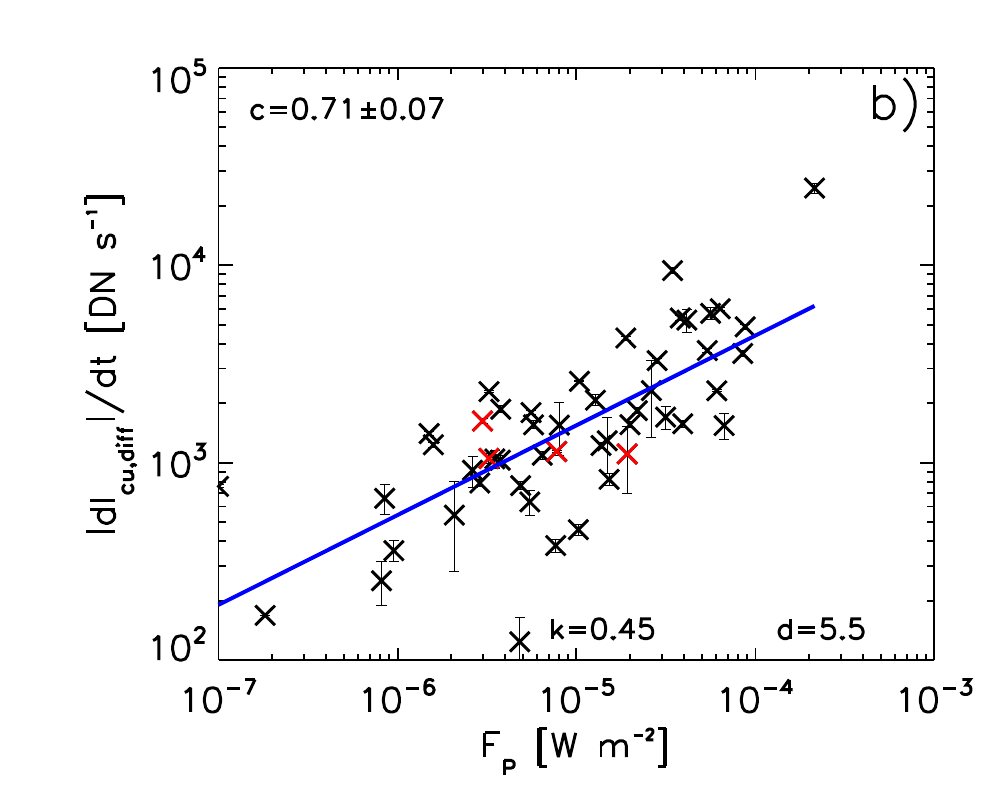}
\caption{Panel (a): The magnetic area $A_{\Phi}$ of the dimming regions against the GOES SXR fluence $F_{T}$. Panel (b): The maximal absolute brightness change rate $|\dot{I}_{\text{cu,diff}}|$ against the GOES SXR peak flux $F_{P}$.
The solid line in blue represents the linear regression to all events in $\log-\log$ space. Red crosses mark event that are not associated with an EUV wave.}
\label{fig:flare_dimming}
\end{figure}

\begin{figure}
\centering
\includegraphics[width=1.0\columnwidth]{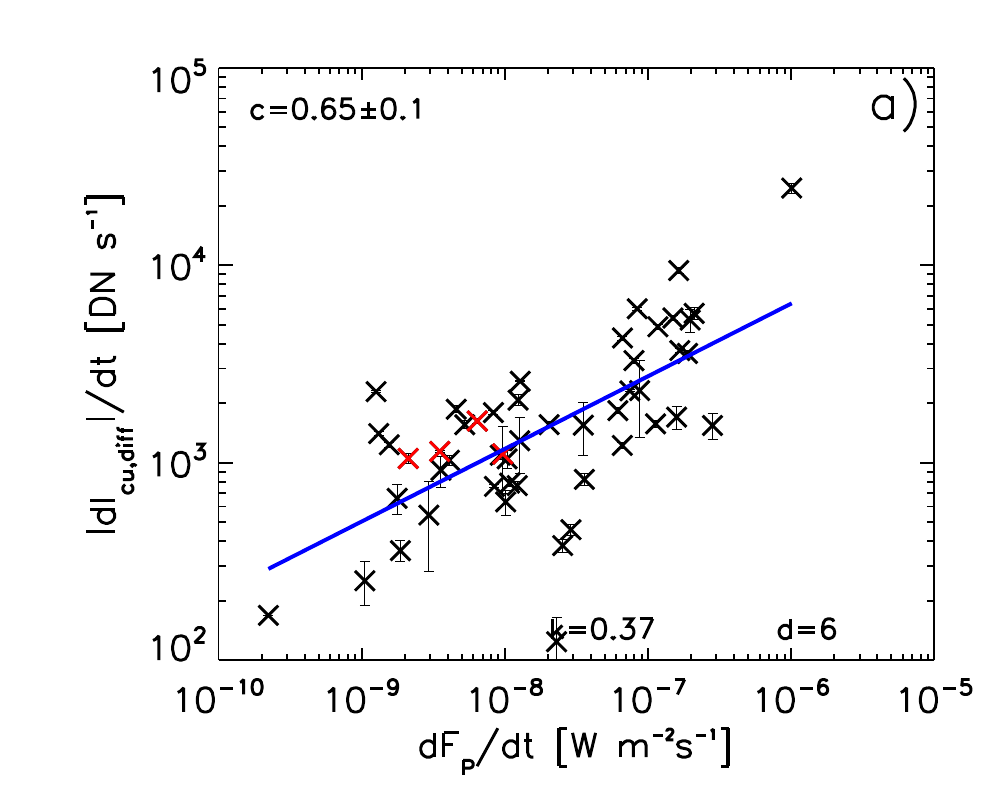}
\includegraphics[width=1.0\columnwidth]{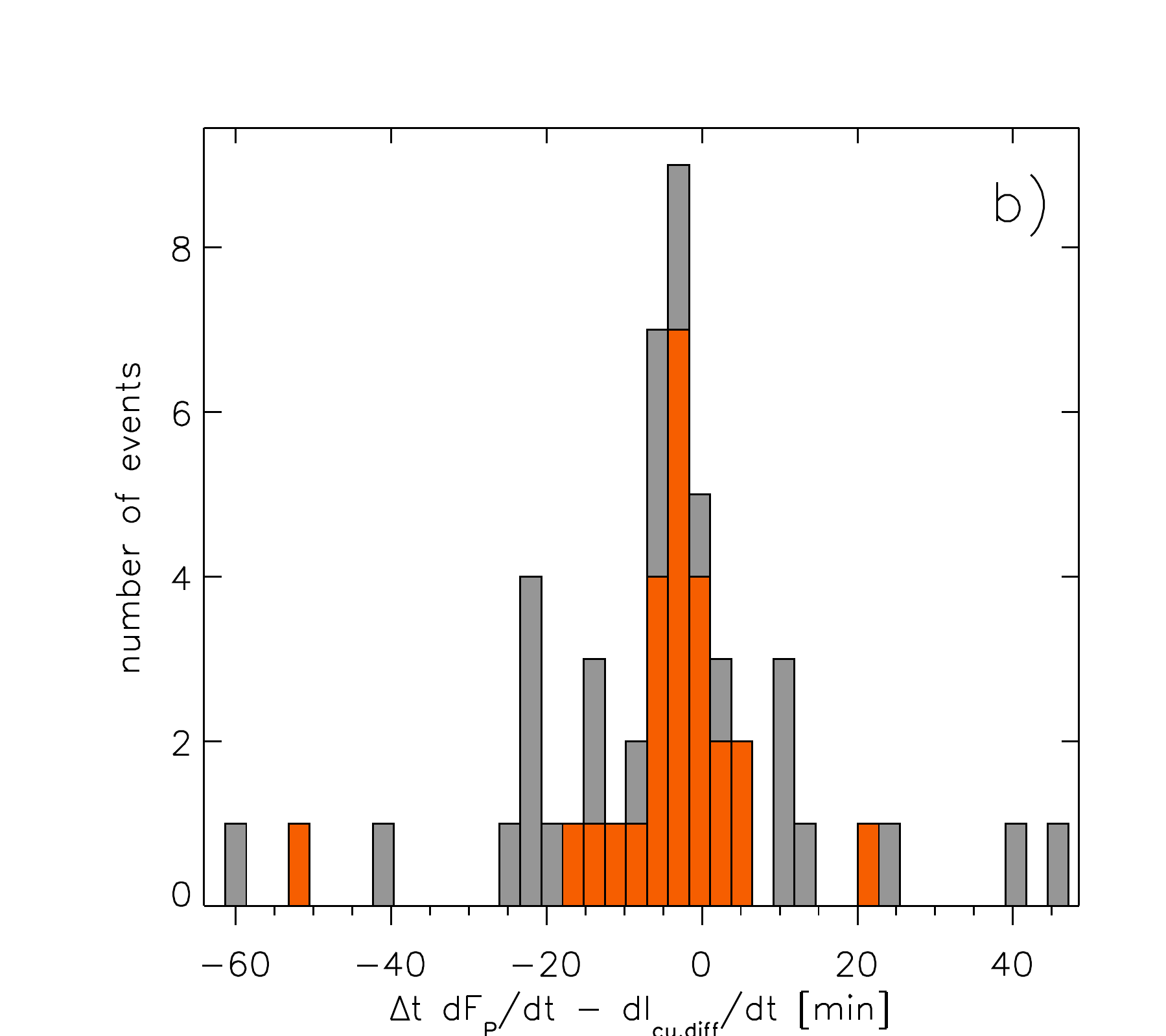}
\caption{Panel (a): The maximal absolute brightness change rate $|\dot{I}_{\text{cu,diff}}|$ against the derivative of the GOES SXR peak flux $\dot{F}_{P}$. Panel (b): Distribution of the time difference between the peak of the GOES SXR derivative $\dot{F}_{P}$ and the minimum in the brightness change rate $\dot{I}_{\text{cu,diff}}$. Events associated with strong flares ($>$M1.0) are overplotted in orange.}
\label{fig:brightness_goes_deriv}
\end{figure}

\begin{figure}
\centering
\includegraphics[width=1.0\columnwidth]
{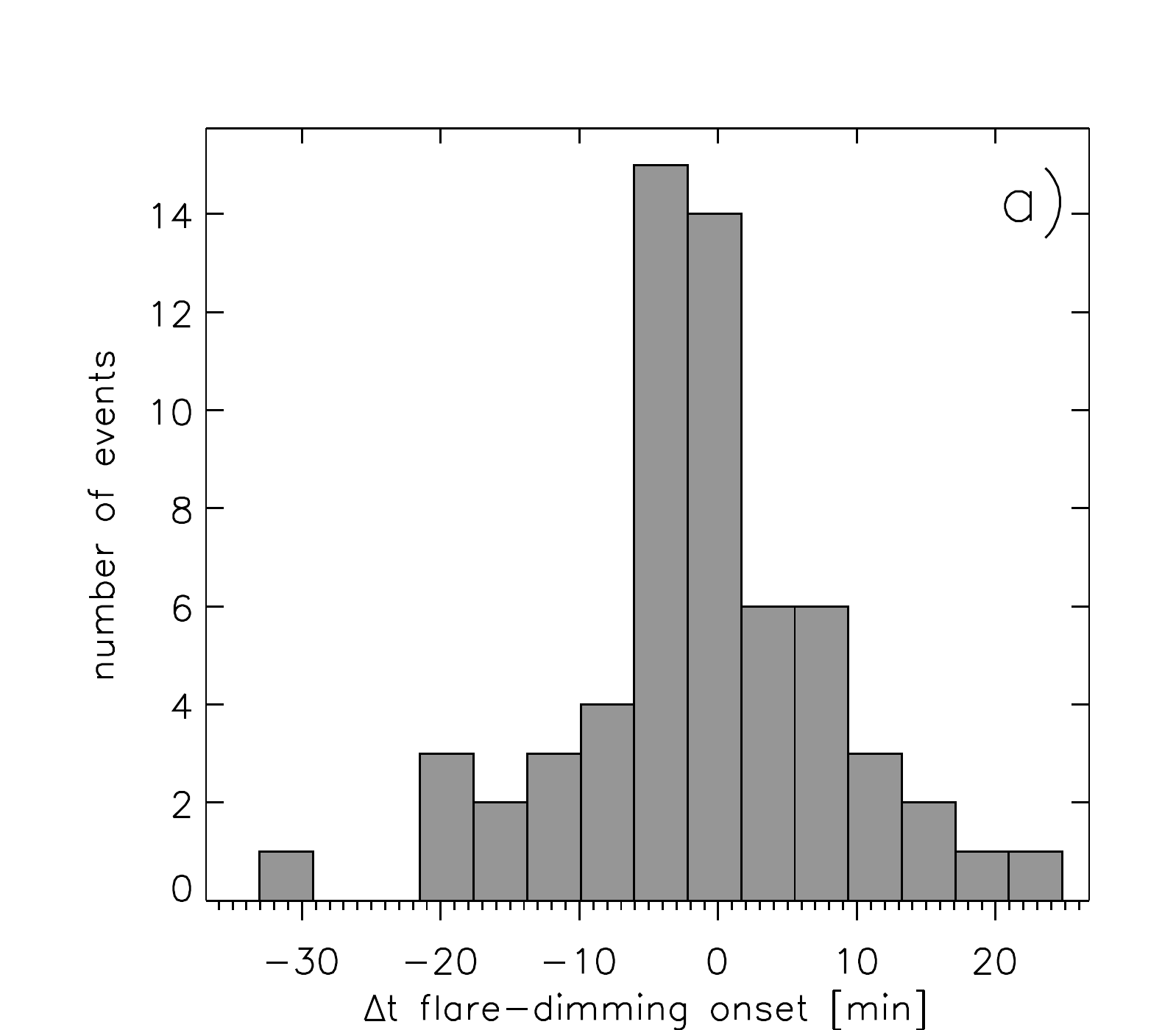}
\includegraphics[width=1.0\columnwidth]{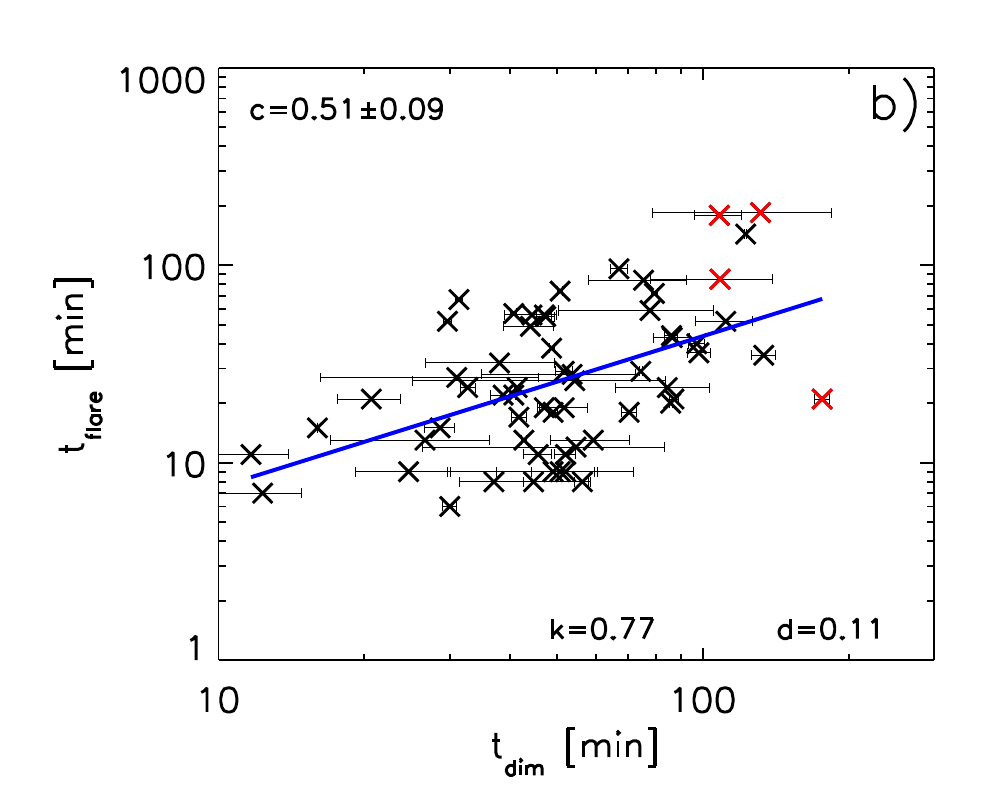}
\caption{Panel (a): Distribution of the time difference between the flare start and the onset of the impulsive phase of the dimming. Panel (b): Correlation plot showing the duration of the flare against the duration of the impulsive phase of the dimming.}
\label{fig:flare_timing}
\end{figure}

\begin{figure}
\centering
\includegraphics[width=1.0\columnwidth]{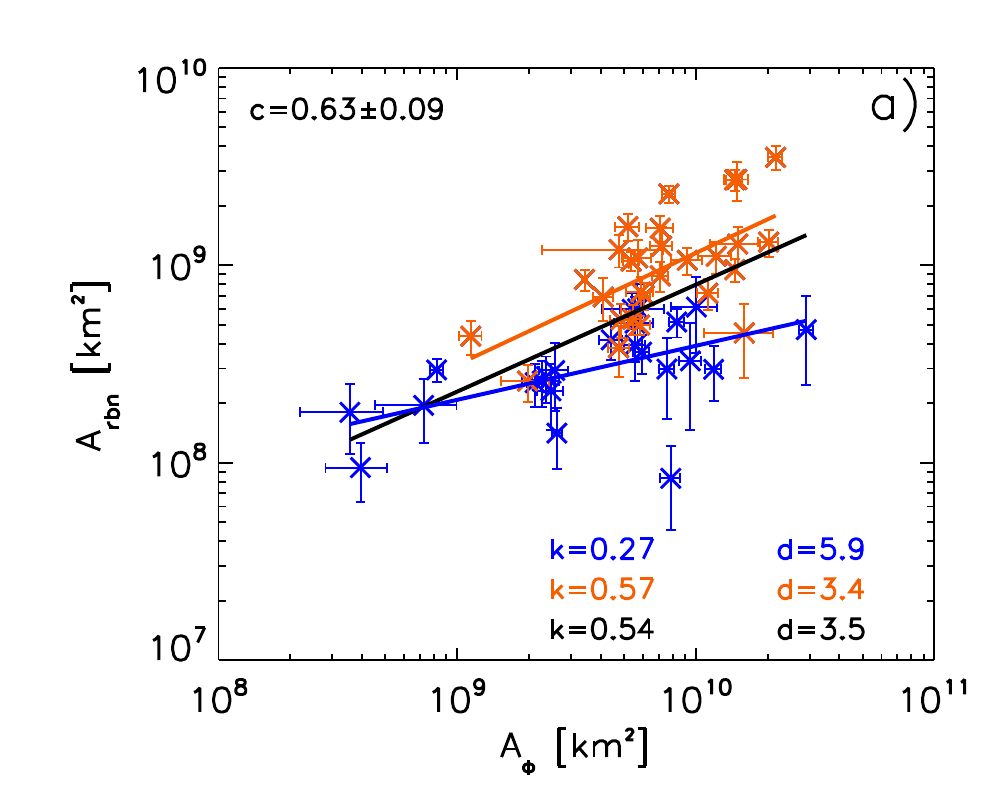}
\includegraphics[width=1.0\columnwidth]{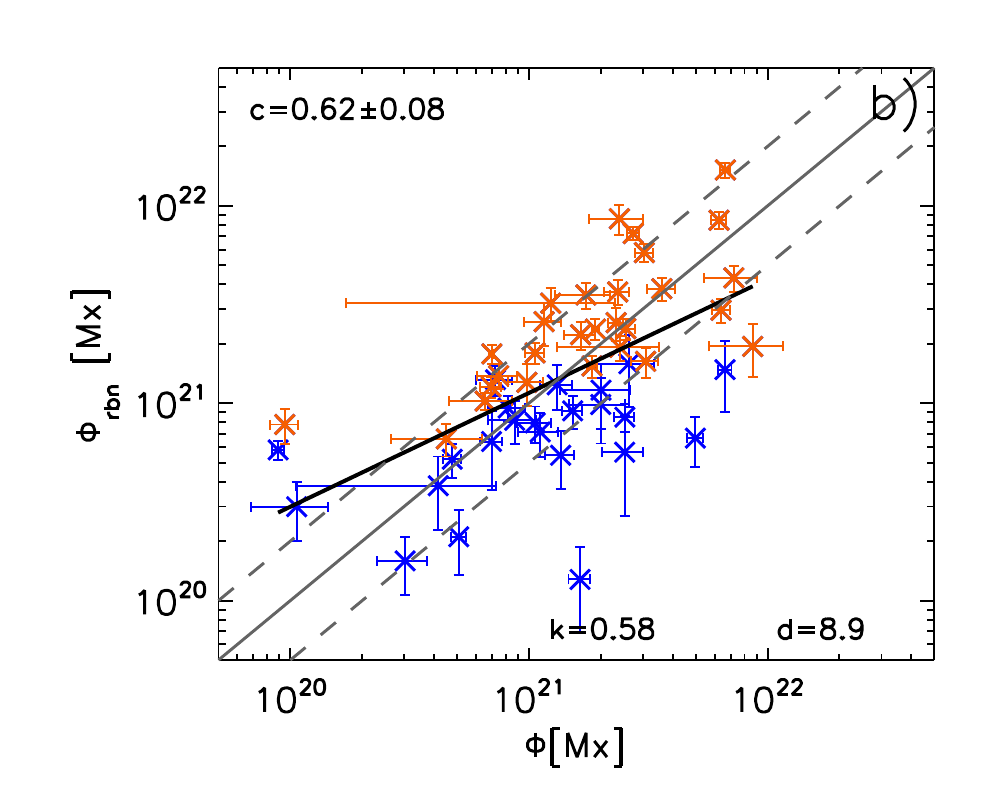}
\caption{Panel (a): Magnetic area $A_{\Phi}$ of the coronal dimmings against the flare ribbon area $A_{\text{rbn}}$. Panel (b): Total unsigned magnetic flux of the dimmings $\Phi$ against the flare ribbon reconnection flux $\Phi_{\text{rbn}}$. Blue crosses mark events that are associated to flares $<$M1.0, orange crosses to $>$M1.0. The regression line is indicated in black. The gray solid and dashed lines represent the 1:1 correspondence line, as well as flux ratios of 0.5 and 2, respectively.}
\label{fig:flare_ribbons}
\end{figure}

In addition, properties of the associated flares obtained from flare ribbon observations, are compared with coronal dimmings. Flare ribbons correspond to the footpoints of newly reconnected flux tubes in the flare arcade in the chromosphere observed in H$\alpha$, (E)UV and hard X-rays. Assuming a direct relationship between the
reconnection flux in the corona and the magnetic flux
swept by the flare ribbons \citep{Forbes:1984}, \cite{Kazachenko:2017} calculated the cumulative magnetic reconnection fluxes of solar flares using flare ribbon observations in the SDO/AIA 1600 \AA~filter. Their event list includes more than 3000 flares stronger than C1.0 that occurred between April 2010 and April 2016. Thus, for 51 events (82\%) of our sample, we could compare the coronal dimming properties with the flare ribbon areas and reconnection fluxes from \cite{Kazachenko:2017}. In Table~\ref{tab:events} we list the flare parameter values for each event. Note that  for the flare reconnection fluxes we list the original values given in \cite{Kazachenko:2017} in Table~\ref{tab:events}, while in order to correctly compare these fluxes with the magnetic fluxes of the dimming regions we divide them by a factor of 2 due their different definitions. Figures \ref{fig:flare_dimming}--\ref{fig:flare_ribbons} show selected, significant correlation plots between dimming and flare parameters together with the linear fits. The correlation matrix given in Figure~\ref{fig:correlations} also includes the correlation coefficients for all possible dimming/flare parameter pairs together with their p-values.

Figure~\ref{fig:flare_dimming}~(a) shows the magnetic area $A_{\Phi}$ of the dimmings against the flare SXR fluence $F_{T}$. The correlation coefficient results in $c=0.72\pm 0.06$, which is the highest correlation coefficient that we find between dimming and flare quantities. A similarly strong correlation is found with the total unsigned magnetic flux $\Phi$ ($c=0.68\pm0.08$), and the absolute minimum dimming brightness ($c=0.69\pm0.08$). This implies, that the higher the flare SXR fluence, the bigger the magnetic dimming area, the higher its magnetic flux and the darker the dimming region.

In Figure~\ref{fig:flare_dimming}~(b) we plot the absolute maximal brightness change rate $\dot{I}_{\text{cu,diff}}$ as a function of the GOES SXR peak flux $F_{P}$. We obtain a correlation coefficient of $c=0.71\pm0.07$, indicating that dimmings that darken faster are associated to stronger flares. $F_{P}$ also shows a strong dependence on the total dimming magnetic flux rate $\dot{\Phi}$ ($c=0.67\pm 0.08$) and on the maximal magnetic area growth rate $\dot{A}_{\Phi}$ ($c=0.56\pm0.1$). Coronal dimmings, that grow and darken fast, tend to be associated with stronger flares.

Figure~\ref{fig:brightness_goes_deriv}~(a) shows the dependence of the absolute maximal brightness change rate $\dot{I}_{\text{cu,diff}}$ with the derivative of the GOES SXR peak flux $\dot{F}_{P}$. The correlation coefficient results in $c=0.65\pm0.1$, i.e.~coronal dimmings associated with flares with higher energy release rates reveal a faster rate of darkening. A positive correlation is also found with the total magnetic flux rate $\dot{\Phi}$ ($c=0.47\pm0.1$) and the maximal magnetic area growth rate $\dot{A_{\Phi}}$ ($c=0.34\pm0.1$).
The temporal relationship of the two parameters is shown in panel~(b), where the distribution of the time difference between the maximum of the GOES SXR derivative and the maximum in the brightness change rate is plotted. Values range from \mbox{$-61.3$ to $44.4$~min}, with a mean of $-5.0\pm18.2$~min and a median of $-3.6\pm11.8$~min. For 80\% of the events associated with strong flares ($>$M1.0), the time difference lies within $\pm 10$~min. We note that the time difference between the peaks of $\dot{F}_{P}$ and $\dot{I}_{\text{cu,diff}}$ is smaller compared to the time difference between the peaks of $\dot{F}_{P}$ and the area growth rate of the dimming $\dot{A}$. This indicates that the combination of intensity decrease and area growth rate in the dynamics of coronal dimmings is closer related to the flare energy release than the area growth rate alone.  

Figure~\ref{fig:flare_timing}~(a) shows the distribution of the time difference between the start of the flare and the onset of the impulsive phase of the dimming. The maximum is reached between $-6.1$ and $-2.2$~min with a mean of $-1.53 \pm 9.9$~min and a median of $-2.0 \pm 7.1$~min. This means that on average the flare onset occurs slightly before the dimming onset.  For 75\% of the events the time difference between the onsets is $|\Delta t|< 10$~min and for almost 50\% it is $< 5$~min. 
Figure~\ref{fig:flare_timing}~(b) presents the correlation plot of the flare duration $t_{\text{flare}}$ and the impulsive phase of the dimming $t_{\text{dim}}$. We find a moderate correlation of $c=0.51 \pm 0.09$, indicating that long duration flare events are associated with long duration dimming events, i.e.~dimming events that show for a long time a significant increase in their area.

\cite{Lin:2004} discuss the possibility that independent of the existence of a flux rope before the eruption, a significant amount of poloidal flux is added to the flux rope during the eruption via magnetic reconnection. In this scenario magnetic fluxes flowing from each end of the current sheet must be equal, since the magnetic flux outside the reconnection site is conserved and the divergence-free condition holds. This implies that the magnetic flux in secondary dimming regions, that represent the footprints of overlying fields that are stretched during the eruption and closed down by subsequent magnetic reconnection, should equal the reconnected flux associated with the flare. In order to test this hypothesis, Figure~\ref{fig:flare_ribbons} shows the relationship between the relevant flare ribbon and dimming parameters. 

Figure~\ref{fig:flare_ribbons}~(a) shows the flare ribbon area $A_{\text{rbn}}$ against the magnetic dimming area $A_{\Phi}$. Since on average the area of core dimming regions make up only $\sim5\%$ of the total dimming, the total dimming regions are also representative for the secondary dimmings. We find a strong correlation of $c=0.63\pm0.09$. Color-coding our data set based on the associated GOES flare class (blue crosses correspond to flare $<$M1.0, orange crosses to $>$M1.0) reveals a clear separation in the distribution and differences in the slope of the regression lines by a factor of 2. 

Figure~\ref{fig:flare_ribbons}~(b) presents the dependence of the flare ribbon reconnection fluxes $\Phi_{\text{rbn}}$ on the total unsigned magnetic flux $\Phi$ of the secondary dimming regions, estimated as $\sim80\%$ of the magnetic flux of the total dimming regions, which is based on the results of the average contribution of the core dimmings derived (see Figure~\ref{fig:hist_core}). A strong correlation coefficient of $c=0.62\pm0.08$ is obtained, indicating that the more magnetic flux is reconnected in the flare, the more magnetic flux is involved in secondary dimmings. Including information on the flare class by color coding events with flares $>$M1.0 in orange and $<$M1.0 in blue shows again a clear separation of the distribution. 
Secondary dimmings associated with strong flares ($>$M1.0) tend to contain roughly the same amount of magnetic flux as the reconnection flux derived from the flare ribbons. This is indicated by the gray solid and dashed lines indicating the identical flux regime. The majority of events of this subset are well distributed within this area. For weaker flares on the other hand, the flare reconnection fluxes are mostly smaller compared to the magnetic fluxes derived from secondary dimming regions.  
\section{Summary and Discussion}
We present a detailed statistical study on characteristic coronal dimming parameters using SDO/AIA and HMI data. Our sample includes 62 on-disk events that allowed the detection of the coronal dimming region, as well as its impulsive phase, uniquely. In the following, we summarize the most important findings:
\begin{enumerate}
\item We identified coronal dimmings in seven EUV filters of SDO/AIA. Channels sensitive to quiet Sun coronal temperatures, such as 211, 193, and \mbox{171 \AA}~were best suited for the dimming detection (211 and 193 \AA: 100\%, 171 \AA: 92\%, 335 \AA: 94\%, 94 \AA: 63\%, 131 \AA: 47\%, 304 \AA: 15\%). Thus, the \mbox{211 \AA}~filter is used for deriving the dimming parameters and for performing the statistical analysis.

\item On average, dimming events reach a size of \mbox{$2.15\times10^{10}$~km$^{2}$} and contain a total unsigned magnetic flux of $1.75\times10^{21}$~Mx (based on the log normal fit to their distributions, see Figure~\ref{fig:hist_area}~(a), \ref{fig:hist_magnetic_flux}~(a)). Both quantities vary over almost three orders of magnitude for the total event sample (see typical ranges in Table~\ref{tab:histograms}). The positive and negative magnetic flux in the dimming region are roughly balanced ($c=0.83\pm0.05$, $k=0.87$, see Figure~\ref{fig:magnetic_flux_balance}), with a mean unsigned magnetic flux density of \mbox{$\sim$ 61~G}.

\item The total brightness of coronal dimmings is on average $-1.91\times10^{6}$~DN (confidence level: $[-5.52,-0.66]\times10^{6}$, see Figure~\ref{fig:hist_brightness}~(a)), which results in a mean relative brightness drop of $\sim$60\% compared to the pre-eruption level. 

\item The duration of the impulsive phase of the dimming, i.e.~the time range where most of the dimming region is evolving, lasts on average \mbox{$\sim$59~min} (median: \mbox{$\sim$50~min}) and for \mbox{$\sim$90\%} of the events it is $<$100~min.
Events that are not associated with an EUV wave are characterized by a long dimming duration ($>$100~min), resulting from a long descend time ($>$60~min) of their impulsive phase (see Figure~\ref{fig:hist_duration}).

\item In 60\% of the events we identify core dimming regions. They contain $\sim$20\% of the total flux but only account for 5\% of the total dimming area (see Figure~\ref{fig:hist_core} \& \ref{fig:magnetic_flux_balance}). The mean relative brightness decrease compared to their pre-eruption intensity level is about 75\%, which means that they are darker compared to the total dimming region (see Figure~\ref{fig:hist_bright_total_core}).

\item The absolute mean brightness of the dimming region $\bar{I}_{\text{cu,diff}}$ and the mean unsigned magnetic flux density $\bar{B}_{\text{us}}$ are positively correlated ($c=0.56\pm0.09$), while $\bar{I}_{\text{cu,diff}}$ does not depend on the size of the dimming region ($c=0.1\pm0.2$). This indicates that the stronger the underlying magnetic flux density, the darker the coronal dimmings are on average. 

\item The flare SXR fluence $F_{T}$ shows a distinct positive correlation with the magnetic area of the dimming $A_{\Phi}$, its total unsigned magnetic flux $\Phi$ and absolute total brightness $|I_{\text{cu,diff}}|$ (all $c\sim 0.7$), while the peak of the SXR flux $F_{P}$ shows the highest correlations with the maximum/minimum of the derivatives of these quantities ($c\sim0.6-0.7$, see Figure~\ref{fig:flare_dimming}).

\item The maximum of the time derivative of the flare SXR flux $\dot{F_{P}}$ correlates with the absolute brightness change rate $|\dot{I}_{\text{cu,diff}}|$ of the dimmings ($c=0.65\pm0.1$, see~Figure~\ref{fig:brightness_goes_deriv}). The total magnetic flux rate $\dot{\Phi}$ ($c=0.47\pm0.1$) and the maximal magnetic area growth rate $\dot{A}_{\Phi}$ ($c=0.34\pm0.1$) also tend to be related to $\dot{F_{P}}$. 

\item Coronal dimmings and flares are closely related in terms of timing: for more than 50\% of the events the time difference between the flare onset and the start of the impulsive phase of the dimming is $|\Delta t|<5$~min, and the mean is $-1.5$~min. In addition, for strong flares ($>$M1.0) the maximum of the GOES SXR derivative and minimum in the brightness change rate of the dimmings occur almost simultaneously. For 80\% of these events, the time difference lies within $\pm$ 10~min (see Figure~\ref{fig:brightness_goes_deriv}). Furthermore, the duration of the impulsive phase of the dimming $t_{\text{dim}}$ and the flare duration $t_{\text{flare}}$ are positively correlated, $c=0.51\pm 0.09$ (see Figure~\ref{fig:flare_timing}).

\item The comparison of flare ribbon properties with dimming parameters revealed a strong positive correlation between the magnetic area of the dimming $A_{\Phi}$ and the flare ribbon area $A_{\text{rbn}}$ (\mbox{$c=0.63\pm0.09$}) as well as the magnetic flux of secondary dimmings $\Phi$ and the flare reconnection fluxes $\Phi_{\text{rbn}}$ with $c=0.62\pm0.08$. In addition, there seems to be a separation in the distribution between stronger ($>$M1.0) and weaker flares ($<$M1.0). 
For strong flares ($>$M1.0), the flare reconnection and secondary dimming fluxes are roughly equal, while for weaker flares the reconnection fluxes are mostly smaller compared to the dimming fluxes (see Figure~\ref{fig:flare_ribbons}). 
\end{enumerate}

The comparison of our results with previous statistical studies is not trivial, since the definition of the parameters describing the properties of coronal dimmings is different and different imaging instruments for the detection were used.
Nevertheless, the typical range of dimming areas found by \cite{Aschwanden:2016} is consistent with our results. 
\cite{Krista:2017} reported a mean dimming size of $4.0\times 10^{9}$~km$^{2}$, which agrees with our findings for the area of core dimming regions. Dimmings within their catalog are analyzed using SDO/AIA 193 \AA~direct images (whereas we used logarithmic base-ratio images). This means that coronal dimmings extracted within this catalog decreased strongly enough in intensity to produce signatures in direct observations, further indicative for core dimmings. In addition, they separated their events in three categories, based on their morphology: single, double and fragmented. We introduced similar classification groups for core dimmings (see Section~\ref{sec:results_core}), implying that regions we identified as core dimmings relate to the dimming regions studied in \cite{Krista:2017}. 

The darker the total brightness of coronal dimmings, the more magnetic flux is involved resulting from higher mean magnetic flux densities. This is in agreement with the finding that core dimmings show a stronger decrease in intensity than total dimming regions and their mean unsigned magnetic flux densities are also stronger. We conclude that localized regions of high density, rooted in regions of strong magnetic flux densities are evacuated and dominate the total brightness of the dimmings.

For the first time, the magnetic properties of the total dimming regions, including the localized core dimmings and the more widespread secondary dimmings, were statistically analyzed in detail. The positive and negative magnetic flux in the total dimming region is balanced and the magnetic flux attributed to secondary dimming regions ($\sim$80\% of the total flux) is strongly correlated with the reconnected flux extracted from the ribbons of the associated flares taken from the \cite{Kazachenko:2017} study ($c=0.62\pm0.08$). These results support the findings by \cite{Lin:2004} that a significant amount of poloidal flux is added to the flux rope during the eruption. 
Therefore, the magnetic flux of secondary dimming regions, representing the overlying expanding fields that are closing down by subsequent magnetic reconnection, should be equal to the reconnected flux estimated from flare ribbons. For flares stronger than M1.0 this is indeed found from our observations: the majority of events shows roughly equal amounts in flare reconnection flux and coronal dimming flux (see Figure~\ref{fig:flare_ribbons}~(b)). 
For weaker flares ($<$M1.0) the magnetic flux involved in secondary dimmings is on average a factor of $\sim2$ higher than the reconnected flux.
We note that the deviation for weaker flares may be related to some systematics (underestimation) in the detections of the UV flare area in weak events (see the deviations in the relation of flare reconnection flux versus GOES class in Figure 8  in \cite{Kazachenko:2017}; whereas \cite{Tschernitz:2018} found a unique relation over the full range of flares from H$\alpha$ data, see their Figure 6).
In addition, stronger flares also show a smaller time difference between the peak of the GOES SXR derivative $\dot{F}_{P}$, and the maximal brightness change rate $\dot{I}_{\text{cu,diff}}$ (see orange distribution in Figure~\ref{fig:brightness_goes_deriv}~(b)).

Regions that we identified as core dimmings are located in opposite polarity regions, which we presume to be locations where flux ropes might be built or already exist. These regions show strong mean magnetic flux densities, and the strongest decrease in intensity in the overall dimming region, indicating that dense plasma previously confined in the low corona is evacuated there. This is in agreement with findings of \cite{Vanninathan:2018}, that showed also localized density drops by up to 70\% in these regions. These regions mark footpoints of the erupting flux rope that most probably existed already before the eruption.

\cite{Temmer:2017} also concluded that the magnetic flux rope of the CME within their study is fed by two components involved in the eruption, namely low-lying magnetic fields (rooted in core dimmings) and sheared overlying magnetic fields (rooted in flare ribbons). In $\sim60\%$ of our events either one or both potential footpoints of this flux rope could be identified, suggestive of an existing flux rope that is erupting (see \cite{Green:2018} for models). 
Furthermore, a number of previous studies also showed a strong relationship between magnetic reconnection and the structure and dynamics of the associated CME/ICME \citep{Qiu:2005,Qiu:2007, Gopalswamy:2017b,Gopalswamy:2017a,Welsch:2017}. 
\cite{Qiu:2005} and \cite{Tschernitz:2018} reported a very strong linear correlation between the reconnection flux swept by flare ribbons and the CME speed ($c\approx0.9$). \cite{Qiu:2007}, \cite{Gopalswamy:2017b}, and \cite{Gopalswamy:2017a} showed that the ribbon fluxes are correlated with the poloidal magnetic flux measured in-situ at 1~au. 

Based on the statistical comparison of dimming and basic flare parameters, we identify first-order dimming parameters, i.e.~parameters reflecting the total extent of the dimming, such as the area, total unsigned magnetic flux and the absolute minimum brightness. This group of parameters show the highest correlation with the SXR fluence of the associated flares. This implies that the larger the SXR fluence of the associated flare, the larger and darker is the dimming region, and the more magnetic flux is involved. Note, the SXR fluence is basically the product of the SXR peak flux and the flare duration. It is a measure of the total flare radiation loss in the \mbox{1--8 \AA}~SXR band, and has been shown to be strongly correlated with the flare energy released \citep{Emslie:2005}.
\cite{Yashiro:2009} found the highest correlations for the mass and the kinetic energy of CMEs with the SXR flare fluence ($c=0.55$ and $c=0.62$, respectively), indicative of a strong relationship between first-order dimming parameters and the CME mass.

Events that are not associated with an EUV wave are preferentially grouped at the high value regime of the distributions of first-order parameters, while they show no significant difference from the original distribution for the corresponding time derivatives. Their flare class does not exceed C-level for most of the cases, however they correspond to long duration events (see Figure~\ref{fig:flare_timing}~(b)), and therefore further confirm the relationship to the SXR flare fluence. 

Furthermore, we identify second-order dimming parameters, as parameters extracted from the derivatives of first-order parameters, such as the maximal area growth rate ($\dot{A}$, $\dot{A}_{\Phi}$), the maximal magnetic flux rate ($\dot{\Phi}$), and the maximal brightness change rate ($\dot{I}_{\text{cu,diff}}$). These parameters describe the dynamics of coronal dimmings. For this class the strongest correlations were found with the peak of the GOES SXR flux $F_{P}$, reflecting the flare strength ($c=0.55\pm0.1$, $c=0.67\pm0.08$, and $c=0.71\pm0.08$, respectively). The stronger the associated flare, the faster the dimming is growing and darkening, and the more magnetic flux is ejected by the CME. 
Positive correlations with the GOES peak flux were also found for the peak velocities of CMEs \citep{Vrsnak:2005,Maricic:2007,Bein:2012}, indicating that these group of dimming parameters may reflect the speeds of the associated CMEs in the low corona.

The time derivative of the GOES SXR flux $\dot{F}_{P}$ is often used as a proxy of the time profile of the flare energy release, according to the Neupert effect that relates the non-thermal and thermal flare emissions \citep[e.g.][]{Dennis:1993, Veronig:2002}. In our study, we found a significant correlation between the brightness change rate of the dimming $\dot{I}_{\text{cu,diff}}$ and the SXR derivative $\dot{F}_{P}$ ($c=0.65\pm0.1$). Also the total magnetic flux rate $\dot{\Phi}$ tends to be related ($c=0.47\pm0.1$), although we note that the correlations with the peak of the SXR flux $F_{P}$ are higher. Several studies show a close temporal relationship between the maximum CME acceleration and the derivative of the flare SXR flux \citep{Zhang:2001,Zhang:2004,Maricic:2007} as well as the RHESSI hard X-ray flux \citep{Temmer:2008,Berkebile-Stoiser:2012}. 
We also find a close temporal relationship between the derivative of the GOES SXR flux and the minimum of the brightness change rate ($-3.6\pm11.8$~min, Figure~\ref{fig:brightness_goes_deriv}~(b)), providing strong evidence that the timing of the flare energy release and the dimming dynamics (brightness change rate) is well synchronized. 

For the majority of events, the formation of coronal dimmings starts $\sim$2~min after the SXR onset of the flare (see Figure~\ref{fig:flare_dimming}), which is in contrast to findings reported in \cite{Maricic:2007} and \cite{Bein:2012}, where the acceleration phase of CMEs starts prior to the flare onset. This difference in timing could be due to the fact that the impulsive phase of the dimming reflects the main phase of plasma evacuation and expansion of the overlying field, which needs some time to be build up to be detected as a coronal dimming signature. Note, that small-scale coronal dimmings may occur prior to the main evacuation phase and therefore could also occur prior to the flare start (see e.g. Figure~\ref{fig:profiles_20110621}).

\section{Acknowledgements}
This study was funded by the Austrian Space Applications Programme of the Austrian Research Promotion Agency FFG (ASAP-11 4900217, BMVIT). A.M.V. also acknowledges the Austrian Science Fund FWF (P24092-N16). SDO data are courtesy of NASA/SDO and the AIA, and HMI science teams. GOES is a joint effort of NASA and the National Oceanic and
Atmospheric Administration (NOAA).

\bibliographystyle{apj}
\bibliography{biblio} 
\end{document}